\documentclass[twocolumn,aps,pra,superscriptaddress]{revtex4-1}

\usepackage{mathpazo}
\usepackage{color}
\usepackage{amsmath}
\usepackage{amssymb}
\usepackage{graphicx}
\usepackage[colorlinks,citecolor=blue,filecolor=black,linkcolor=red,
urlcolor=blue]{hyperref}
\usepackage{breakurl}

\begin{document}

\title{Zero-dimensional limit of the two-dimensional Lugiato-Lefever equation}
\author{Wesley B. Cardoso}
\affiliation{Instituto de F\'isica, Universidade Federal de Goi\'as, 74.690-900,
Goi\^ania, Goi\'as, Brazil}
\author{Luca Salasnich}
\affiliation{Dipartimento di Fisica e Astronomia ``Galileo Galilei'' and CNISM,
Universit\`a di Padova, Via Marzolo 8, 35131 Padova, Italy}
\affiliation{
Istituto Nazionale di Ottica (INO) del Consiglio Nazionale delle Ricerche
(CNR), Sezione di Sesto Fiorentino, Via Nello Carrara, 1 - 50019 Sesto
Fiorentino, Italy}
\author{Boris A. Malomed}
\affiliation{Department of Interdisciplinary Studies, School of Electrical Engineering,
Faculty of Engineering, Tel Aviv University, Tel Aviv 69978, Israel}
\affiliation{Laboratory of Nonlinear-Optical Informatics, ITMO University, St. Petersburg
197101, Russia}
%\date{\today }

\begin{abstract}
We study effects of tight harmonic-oscillator confinement on the
electromagnetic field in a laser cavity by solving the two-dimensional
Lugiato-Lefever (2D LL) equation, taking into account self-focusing or
defocusing nonlinearity, losses, pump, and the trapping potential. Tightly
confined (quasi-zero-dimensional) optical modes (\textit{pixels}), produced
by this model, are analyzed by means of the variational approximation, which
provides a qualitative picture of the ensuing phenomena. This is followed by
systematic simulations of the time-dependent 2D LL equation, which reveal
the shape, stability, and dynamical behavior of the resulting localized
patterns. In this way, we produce stability diagrams for the expected
pixels. Then, we consider the LL model with the vortical pump, showing that
it can produce stable pixels with embedded vorticity (\textit{vortex solitons%
}) in remarkably broad stability areas. Alongside confined vortices with the
simple single-ring structure, in the latter case the LL model gives rise to
stable multi-ring states, with a spiral phase field. In addition to the
numerical results, a qualitatively correct description of the vortex
solitons is provided by the Thomas-Fermi approximation.
\end{abstract}

\maketitle

\section{Introduction}

The Lugiato-Lefever (LL) equation \cite{LL} in one and two dimensions (1D\
and 2D) is a fundamental model governing the dynamics of optical fields in
pumped lossy laser cavities with the intrinsic Kerr nonlinearity, which may
have self-focusing or defocusing sign. This equation is well known as an
important tool for the analysis of pattern formation, with various
applications in nonlinear optics \cite{patterns-early, pixel}. The progress
in theoretical and experimental studies has recently drawn a great deal of
renewed interest to the use of the LL equation in diverse settings \cite%
{patterns-2}-\cite{NJP}. A natural extension of these studies is
incorporation of external potentials into the LL equation, which can be
easily fabricated in laser cavities as transverse landscapes of the
refractive-index inhomogeneity, and may be used as an efficient means for
the control of optical fields \cite{Kestas}.

One of essential applications of the LL equation is its use for modeling
well-localized pixels (i.e., sharply bounded bright spots) in the cavity
\cite{pixel}. In most cases, pixels are considered as \textit{anti-dark
solitons}, i.e., bright objects created on top of a uniformly pumped
background field. In this work, we aim to demonstrate a possibility to
create completely localized robust pixels (i.e., bright solitons with zero
background), by adding to the model a confining potential corresponding to
an isotropic 2D harmonic oscillator. Furthermore, we demonstrate that the
same setting makes it possible to create stable \textit{vortex pixels}, by
applying a vortically structured pump. The consideration reported below
combines an analytical approach, chiefly based on the variational and
Thomas-Fermi approximations (VA and TFA), and systematic direct simulations,
in imaginary and real time alike, with the purpose to create confined modes
and test their stability.

The paper is organized as follows. The model, based on the 2D\ LL equation
with the harmonic-oscillator trapping potential, is formulated in Section
II. Analytical treatment, which makes use of the VA, power-balance equation,
and TFA, is presented in Section III. Numerical results for the existence
and stability of the fundamental (zero-vorticity) and vortical trapped modes
are reported in Sections IV and V, respectively. The latter section also
reports simple analytical results for the vortex states, obtained by means
of the TFA. The paper is concluded by Section VI.

\section{The model}

The 2D LL equation for the amplitude $\phi (x,y,t)$ of the electromagnetic
field in a pumped lossy laser cavity is (see, e.g., Ref. \cite{Kestas})
\begin{gather}
i\left( \gamma +\frac{\partial }{\partial t}\right) \phi =\left[ -\frac{1}{2}%
\left( \frac{\partial ^{2}}{\partial x^{2}}+\frac{\partial ^{2}}{\partial
y^{2}}\right) +\Delta \right.  \notag \\
+\left. \frac{\Omega ^{2}}{2}(x^{2}+y^{2})+\sigma |\phi |^{2}\right] \phi
+E\;,  \label{LuLe}
\end{gather}%
where $E$ is the pump field, $\gamma >0$ the dissipation rate, $\Delta
\gtrless 0$ detuning of the pump with respect to the cavity, and $\Omega
^{2} $ the strength of the confining potential, while $\sigma =-1$ and $+1$
correspond to the self-focusing and defocusing nonlinearity, respectively.
By means of rescaling, one may fix $\gamma =1$, although it may be convenient
to keep $\gamma $ as a free parameter, as shown below.

Stationary solutions to Eq. (\ref{LuLe}) have a simple asymptotic form at $%
r\equiv \sqrt{x^{2}+y^{2}}\rightarrow \infty $:%
\begin{equation}
\phi (r)\approx -\frac{2E}{\left( \Omega r\right) ^{2}}+\frac{4\left( \Delta
-i\gamma \right) E}{\left( \Omega r\right) ^{4}}.  \label{infi}
\end{equation}%
We also note that the following exact power-balance equation ensues from Eq.
(\ref{LuLe}):
\begin{equation}
\frac{dP}{dt}=-2\gamma P-2\int \int \mathrm{Im}\{E^{\ast }\phi (x,y,t)\}dxdy,
\label{PB}
\end{equation}%
where power $P$ (alias norm) of the solitary wave is defined as
\begin{equation}
P=\int \int |\phi (x,y,t)|^{2}dxdy.  \label{Norm}
\end{equation}

The objective is to reduce the 2D LL equation (\ref{LuLe}) to a
quasi-zero-dimensional limit (a dynamical system for a pixel, similar to
those realized by theoretically predicted \cite{pixel} and experimentally
created \cite{NJP} spatial solitons) in the case of tight confinement,
represented by large $\Omega ^{2}$. First, we do it by means of the VA,
defining
\begin{equation}
\phi (x,y,t)\equiv \Phi (x,y,t)\exp \left( -\gamma t\right) ,
\end{equation}
and thus casting Eq. (\ref{LuLe}) in the form of
\begin{eqnarray}
i\frac{\partial }{\partial t}\Phi &=&\left[ -\frac{1}{2}\left( \frac{%
\partial ^{2}}{\partial x^{2}}+\frac{\partial ^{2}}{\partial y^{2}}\right)
+\Delta \right.  \notag \\
&+&\left. \frac{\Omega ^{2}}{2}(x^{2}+y^{2})+\sigma e^{-2\gamma t}|\Phi |^{2}%
\right] \Phi +Ee^{\gamma t}.  \label{PhiLL}
\end{eqnarray}%
Unlike the original LL equation (\ref{LuLe}), the transformed one (\ref%
{PhiLL}) can be directly derived from a real time-dependent Lagrangian,
\begin{eqnarray}
L &=&\int \int dxdy\left\{ \frac{i}{2}\left( \Phi _{t}^{\ast }\Phi -\Phi
^{\ast }\Phi _{t}\right) +\frac{1}{2}\left( |\Phi _{x}|^{2}+|\Phi
_{y}|^{2}\right) \right.  \notag \\
&+&\left[ \Delta +\frac{\Omega ^{2}}{2}(x^{2}+y^{2})\right] |\Phi |^{2}+%
\frac{\sigma }{2}e^{-2\gamma t}\,|\Phi |^{4}  \notag \\
&+&\left. e^{\gamma t}\left( E\Phi ^{\ast }+E^{\ast }\Phi \right) \right\}
\,.  \label{LLL}
\end{eqnarray}

\section{Analytical considerations}

\subsection{The variational approximation}

For the 1D LL equation without trapping potentials, the VA was developed in
Ref. \cite{VA-LL}. To derive this approximation in a form appropriate for
the present model, we note that, in the lowest approximation, Eq. (\ref%
{PhiLL}) gives rise to the following asymptotic form of solutions at $%
r\rightarrow \infty $: $\Phi =-2E\left( \Omega r\right) ^{-2}e^{\gamma t}$,
cf. Eq. (\ref{infi}). This form suggests us to adopt an ansatz based on the
fractional expression, with real variables $f(t)$ and $\chi (t)$, which may
be combined into a complex one, $F(t)=f(t)\exp \left( i\chi (t)\right) $:%
\begin{eqnarray}
\Phi \left( x,y,t\right) &=&-\frac{2E}{\Omega ^{2}}e^{\gamma t}\frac{F(t)}{%
1+r^{2}F(t)}\equiv \epsilon \,e^{\gamma t}\frac{f(t)\,e^{i\chi (t)}}{%
1+r^{2}f(t)\,e^{i\chi (t)}}\;,\quad  \label{ansatz} \\
\epsilon &\equiv &-\frac{2E}{\Omega ^{2}}\;.  \label{epsilon}
\end{eqnarray}

\begin{figure}[tb]
\includegraphics[width=0.75\columnwidth]{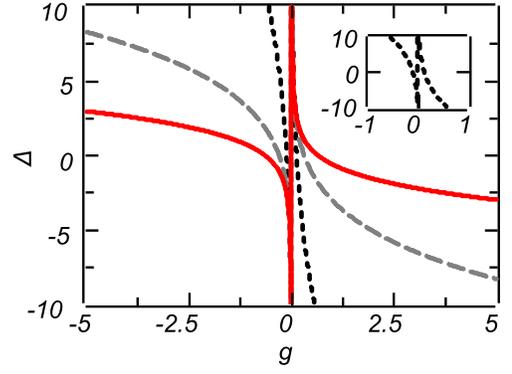}
\caption{(Color online) Lines in the parameter plane of ($\Delta $, $g$),
along which solutions of the overdetermined system (\protect\ref{n21}), (%
\protect\ref{n22}) exist. Here, solid red, dashed gray, and dotted black
lines correspond, respectively, to $\Omega =2$, $4$, and $6$. The inset
displays a zoom of the curve for $\Omega =10$.}
\label{ER}
\end{figure}

The insertion of ansatz (\ref{ansatz}) in Eq. (\ref{LLL}) and subsequent
integration gives rise to an effective Lagrangian,
\begin{eqnarray}
\frac{e^{-2\gamma t}}{\pi \epsilon ^{2}}L_{\mathrm{eff}} &=&\frac{1}{2}%
fq_{1}(\chi )\frac{d\chi }{dt}-\frac{1}{2}q_{2}(\chi )\sin \chi \frac{df}{dt}%
+f^{2}q_{2}(\chi )  \notag \\
&+&\Delta fq_{1}(\chi )+\frac{\sigma \epsilon ^{2}}{8}f^{3}q_{3}(\chi
)-\Omega ^{2}q_{1}(\chi )\cos \chi  \notag \\
&-&\frac{\Omega ^{2}}{4}\int d\chi \lbrack q_{3}(\chi )\sin \chi ]\,,
\label{L}
\end{eqnarray}%
with $q_{1}(\chi )\equiv \chi /\sin \chi $, $q_{2}(\chi )\equiv \lbrack
\left( \sin \chi -\chi \cos \chi \right) /\sin ^{3}\chi $, and $q_{3}(\chi
)\equiv \lbrack 2\chi -\sin \left( 2\chi \right) ]/\sin ^{3}\chi $. The last
term in Eq. (\ref{L}) is cast in the integral form as a result of
``renormalization": the respective term in the original
Lagrangian formally diverges logarithmically at $R\rightarrow \infty $, but
the diverging part actually does not depend on $f$ and $\chi $, and it may
be cancelled by means of the differentiation with respect to $\chi $ and
subsequent integration, also with respect to $\chi $.

The Euler-Lagrange equations following from Lagrangian (\ref{L}) are (taking
into account that the Lagrangian must be substituted into the action, $\int
Ldt$, and then the action must be subjected to the variation; this makes it
necessary to apply the time differentiation to $e^{2\gamma t}$):
\begin{eqnarray}
&&\frac{1}{2}\left[ q_{2}(\chi )\cos \chi +q_{2}^{\prime }(\chi )\sin \chi
+q_{1}(\chi )\right] \frac{df}{dt}  \notag \\
&+&\left( \gamma f-\Omega ^{2}\sin \chi \right) q_{1}(\chi )+\left( \Omega
^{2}\cos \chi -\Delta \,f\right) q_{1}^{\prime }(\chi )  \notag \\
&-&f^{2}\,q_{2}^{\prime }(\chi )-\frac{g}{8}f^{3}q_{3}^{\prime }(\chi )+%
\frac{\Omega ^{2}}{4}q_{3}(\chi )\sin \chi =0\;,  \label{f0}
\end{eqnarray}%
\begin{eqnarray}
&&\Delta q_{1}(\chi )+2\,f\,q_{2}(\chi )+\frac{3g}{8}\,f^{2}\,q_{3}(\chi
)+\gamma q_{2}(\chi )\sin \chi  \notag \\
&+&\frac{1}{2}\left[ q_{2}(\chi )\cos \chi +q_{2}^{\prime }(\chi )\sin \chi
+q_{1}(\chi )\right] \frac{d\chi }{dt}=0\;,  \label{theta0}
\end{eqnarray}%
where a renormalized nonlinearity coefficient is [see Eq. (\ref{epsilon})]
\begin{equation}
g=\sigma \epsilon ^{2}\equiv 4\sigma E^{2}/\Omega ^{4}.  \label{g}
\end{equation}%
Note that, although it may seem that Eqs. (\ref{f0}) and (\ref{theta0}) are
singular at $\chi =0$, in reality all the singularities cancel. A
singularity is instead possible at $\chi =\pi $.

We consider stationary (fixed-point) solutions of Eqs. (\ref{f0}) and (\ref%
{theta0}) by setting $df/dt=d\chi /dt=0$, which yields
\begin{eqnarray}
&&\left( \Omega ^{2}\sin \chi -\gamma f\right) q_{1}(\chi )+\left( \Delta
\,f-\Omega ^{2}\cos \chi \right) q_{1}^{\prime }(\chi )  \notag \\
&+&f^{2}\,q_{2}^{\prime }(\chi )+\frac{g}{8}f^{3}q_{3}^{\prime }(\chi )-%
\frac{\Omega ^{2}}{4}q_{3}(\chi )\sin \chi =0,  \label{f}
\end{eqnarray}%
\begin{equation}
\Delta q_{1}(\chi )+2\,f\,q_{2}(\chi )+\frac{3g}{8}\,f^{2}\,q_{3}(\chi
)+\gamma q_{2}(\chi )\sin \chi =0.  \label{theta}
\end{equation}%
Further, it is possible to find approximate solutions of Eqs. (\ref{f}) and (%
\ref{theta}), assuming that they have
\begin{equation}
|\chi |\ll \pi ~.  \label{<<}
\end{equation}%
In this case, Eq. (\ref{theta}), in the first approximation, assumes the
form of
\begin{equation}
\Delta +\frac{2}{3}f+\frac{g}{2}f^{2}+\frac{\gamma }{3}\chi =0.
\label{new_theta}
\end{equation}%
{\small \ }Similarly, in the lowest approximation Eq. (\ref{f}) yields an
expression for $\chi $:
\begin{equation}
\chi =\frac{30\gamma f}{10\Omega ^{2}+f\left[ 10\Delta +8f+3gf^{2}\right] }%
\;,  \label{th}
\end{equation}%
The assumption (\ref{<<}) may be then secured by a natural assumption of the
strong confinement, i.e., considering large values of $\Omega $. In this
case, Eqs. (\ref{new_theta}) and (\ref{th}) can be further simplified to
\begin{eqnarray}
f &\approx &\frac{-2\pm \sqrt{4-18g\Delta }}{3g},  \label{ff-simple} \\
\chi &\approx &\frac{\gamma \left( -2\pm \sqrt{4-18g\Delta }\right) }{%
g\Omega ^{2}}.  \label{th-simple}
\end{eqnarray}%
Obviously, Eqs. (\ref{ff-simple}) and (\ref{th-simple}) produce a physically
relevant result under condition $g\Delta <2/9$.

One can construct another approximate solution for large detuning $\Delta $:
\begin{eqnarray}
f &\approx &\sqrt{-2\Delta /g}-2/\left( 3g\right) ,  \label{f_DLarge} \\
\chi &\approx &(15/2)\gamma /\Delta .  \label{theta_DLarge}
\end{eqnarray}%
In the general case, stationary solutions of Eqs. (\ref{f}) and (\ref{theta}%
), where, as said above, we may fix $\gamma =1$, depend on three parameters:
$\Delta \gtrless 0$, $g\gtrless 0$ [see Eq. (\ref{g})], and $\Omega ^{2}>0$.

\begin{figure}[tb]
\includegraphics[width=0.48\columnwidth]{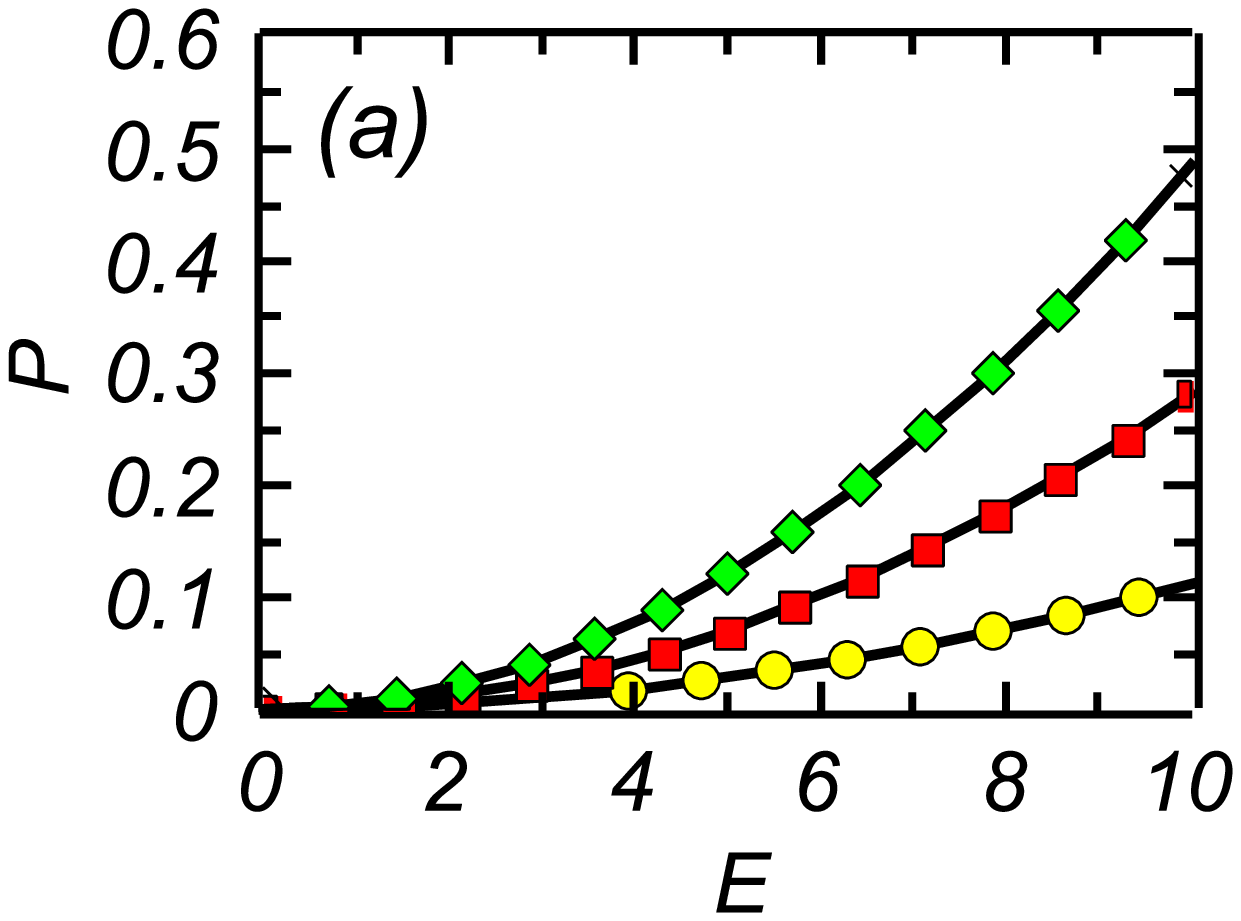} \hfil
\includegraphics[width=0.48\columnwidth]{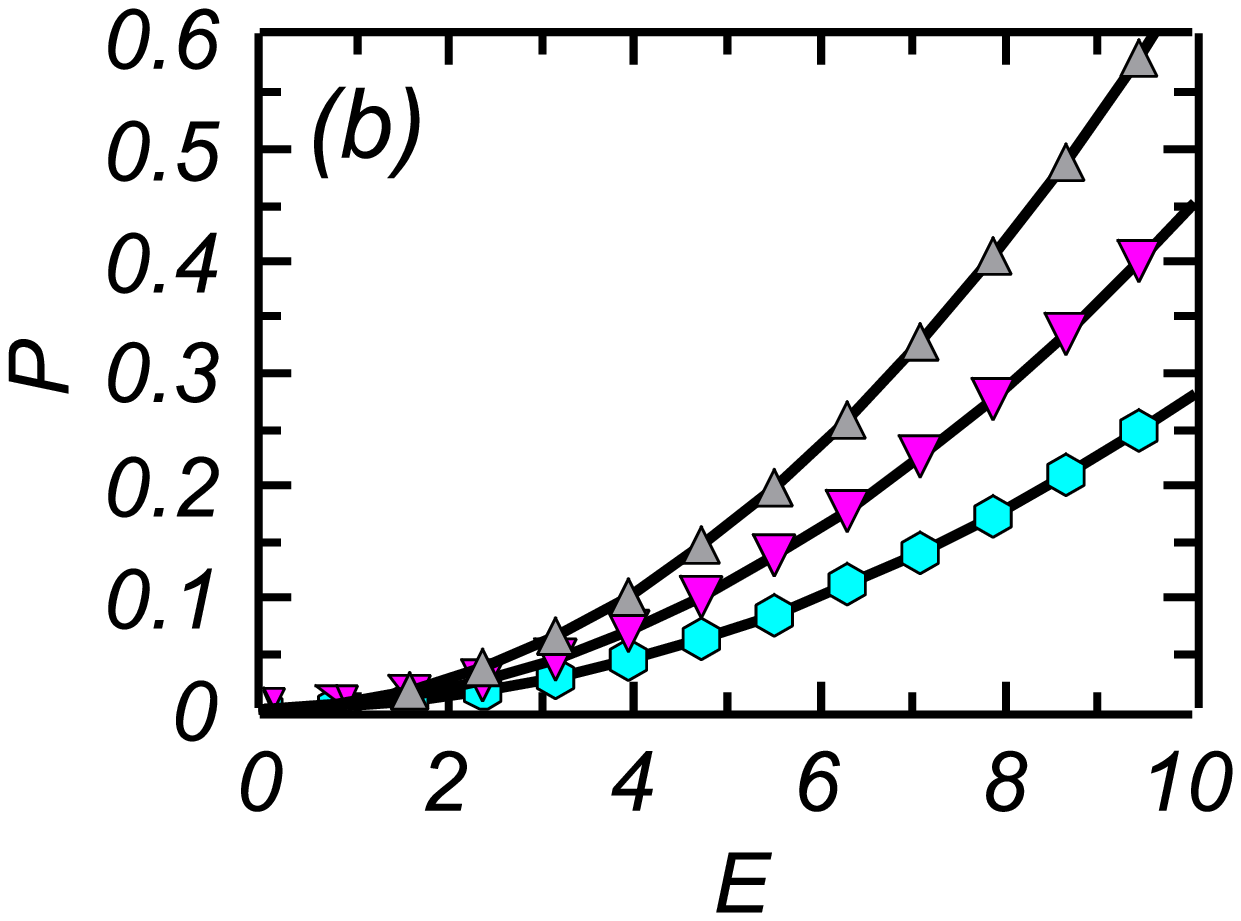} \hfil
\includegraphics[width=0.48\columnwidth]{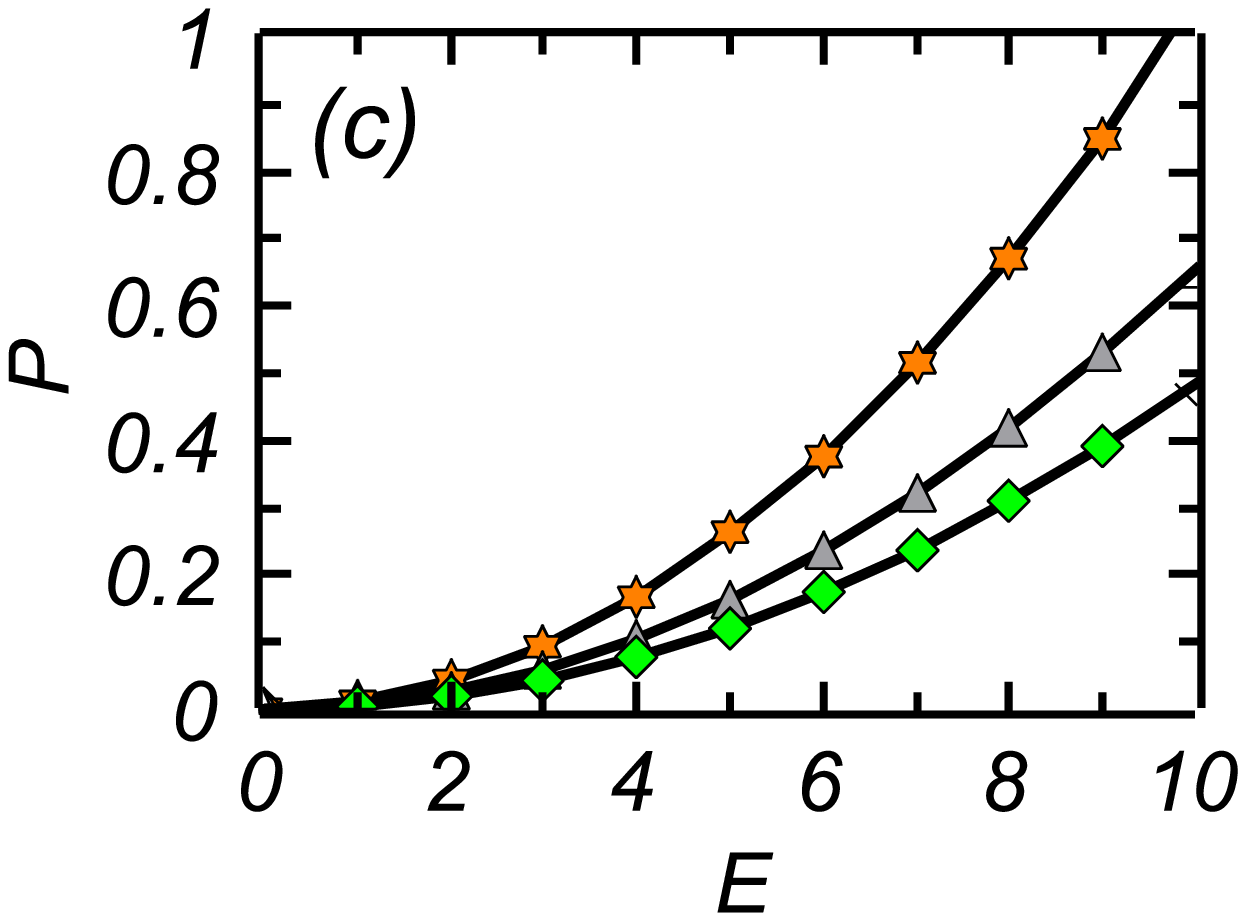}
\caption{(Color online) Power $P$ versus pumping strength $E$. Variational
results for the defocusing case ($\protect\sigma =+1$), produced by
simplified equations (\protect\ref{ff-simple}) and (\protect\ref{th-simple}%
), are shown in (a), with $g=1$, by curves with circles (yellow), boxes
(red), and diamonds (green), for $\Delta =-1$, $-4$, and $-10$,
respectively. The self-focusing case ($\protect\sigma =-1$) is shown in (b),
with $g=-1$, by curves with hexagons (cyan), down triangles (magenta), and
up triangles (gray), for $\Delta =1$, $4$, and $10$, respectively. In (c) we
compare the analytical and numerical results for the defocusing case ($g=1$
and $\Delta =-10$), shown, severally, by curves with diamonds (green) and
stars (orange), and analytical results for the self-focusing case [$g=-1$
and $\Delta =10$, shown by the curve with up triangles (gray)]. The
solutions numerically found in the case of the self-focusing are unstable.
In all the plots, $\Omega =10$ and $\protect\gamma =1$ are fixed here.}
\label{F5}
\end{figure}

In addition to the consideration of the stationary solutions (fixed points),
the full dynamical version of the VA, based on Eqs. (\ref{f0}) and (\ref%
{theta0}), can be also used to analyze their stability, as well as evolution
of unstable solutions. However, in practical terms such a dynamical analysis
turns out to be quite cumbersome, direct numerical simulations being
actually more efficient, as shown below.

\subsection{The power-balance condition}

The substitution of ansatz (\ref{ansatz}) in the definition of power (\ref%
{Norm}) and power-balance equation (\ref{PB}) yields
\begin{eqnarray}
P &=&\frac{4\pi E^{2}f}{\Omega ^{4}}\frac{\chi }{\sin (\chi )},  \label{P_2}
\\
\frac{dP}{dt} &=&-\frac{8\pi \gamma E^{2}f}{\Omega ^{4}}\frac{\chi }{\sin
(\chi )}+\frac{4\pi E^{2}}{\Omega ^{2}}\chi ,  \label{dP_2}
\end{eqnarray}%
(in these expressions, $f>0$ is implied). Equation (\ref{dP_2}) predicts the
equilibrium condition, $dP/dt=0$, at
\begin{equation}
\sin (\chi )=\frac{2\gamma }{\Omega ^{2}}f.  \label{equilibrium}
\end{equation}%
Note that $E$ drops out from Eq. (\ref{equilibrium}), and condition $\sin
(\chi )\leq 1$, following from Eq. (\ref{equilibrium}), imposes a
restriction on $f$,
\begin{equation}
f\leq \Omega ^{2}/(2\gamma ).
\end{equation}%
Finally, for $|\chi |\ll \pi $ Eq. (\ref{equilibrium}) simplifies to $\chi
\approx \left( 2\gamma /\Omega ^{2}\right) f$. Using this to eliminate $f$
in favor of $\chi $, Eqs. (\ref{new_theta}) and (\ref{th}) give rise to the
following system of equations:
\begin{eqnarray}
\frac{g\Omega ^{4}\chi ^{3}}{80\gamma ^{3}}+\frac{\Omega ^{2}\chi ^{2}}{%
15\gamma ^{2}}+\frac{\Delta \chi }{6\gamma }-\frac{1}{6} &=&0,  \label{n21}
\\
\frac{g\Omega ^{4}\chi ^{2}}{8\gamma ^{2}}+\frac{1}{3\gamma }\left( \Omega
^{2}+\gamma ^{2}\right) \chi +\Delta &=&0.  \label{n22}
\end{eqnarray}%
Of course, the system of two equations (\ref{n21}) and (\ref{n22}) for the
single unknown $\chi $ is overdetermined, and a solution of this system may
exist only if a special restriction is imposed on parameters, as shown in
Fig. \ref{ER}, in the plane of ($\Delta $, $g$), for $\gamma =1$ and three
different fixed values of the confinement strength, $\Omega =2$, $\Omega =4$%
, and $\Omega =10$. Note that these curves do not depend on the pumping
strength, $E$. Indeed, this parameter is related only to the power of the
solution, see Eq. (\ref{P_2}) and Fig. \ref{F5}. The meaning of the
overdetermined system is that, realizing the VA and power-balance condition
simultaneously, its solution has a better chance to produce an accurate
approximation. This expectation is qualitatively corroborated below, see
Fig. \ref{NER} and related text in the next section.

Generic properties of the modes predicted by ansatz (\ref{ansatz}) are
characterized by the corresponding dependence of power $P$ on the pumping
strength, $E$. Using, for this purpose, the simplified approximation given
by Eqs. (\ref{ff-simple}) and (\ref{th-simple}), we display the dependences
for the defocusing nonlinearity ($g=1$) in Fig. \ref{F5}(a), at fixed values
of the detuning, $\Delta =-1$, $-4$, and $-10$. Figure \ref{F5}(b) displays
the same dependences in the case of the self-focusing nonlinearity ($g=-1$),
for $\Delta =1$, $4$, and $10$. Note that power $P$ is not symmetric with
respect to the reversal of the signs of nonlinearity $g$ and detuning $%
\Delta $.

In Fig. \ref{F5}(c) we compare the VA\ results for the self-defocusing ($g=1$
and $\Delta =-10$) and focusing ($g=-1$ and $\Delta =10$) cases. In
addition, Fig. \ref{F5}(c) includes full numerical results (for details see
the next Section). It is seen that the simplified VA produces a qualitatively
correct prediction, which is not quite accurate quantitatively. Below, we
demonstrate that the VA is completely accurate only in small black regions
shown in Fig. \ref{NER}.

\subsection{The Thomas-Fermi approximation (TFA)}

In the case of the self-defocusing sign of the nonlinearity, and positive
mismatch, $\Delta >0$, the ground state, corresponding to a sufficiently
smooth stationary solution of Eq. (\ref{LuLe}), $\phi =\phi (r)$, may be
produced by the TFA, which neglects derivatives in the radial equation \cite%
{TFA}:%
\begin{equation}
\left( \Delta -i\gamma +\frac{\Omega ^{2}}{2}r^{2}+\sigma |\phi |^{2}\right)
\phi =-E\;.  \label{TF}
\end{equation}%
In particular, the TFA is relevant if $\Delta $ is large enough.

The TFA-produced equation (\ref{TF}) for the ground-states's configuration
is not easy to solve analytically, as it is a cubic algebraic equation with
complex coefficients. The situation greatly simplifies in the limit case of
a very large mismatch, \textit{viz}., $\Delta \gg \gamma $ and $\Delta
^{3}\gg \sigma E^{2}$. Then, both the imaginary and nonlinear terms may be
neglected in Eq. (\ref{TF}), to yield%
\begin{equation}
\phi (r)\approx -E\left( \Delta +\frac{\Omega ^{2}}{2}r^{2}\right) ^{-1}.
\label{simple}
\end{equation}%
This simple approximation, which may be considered as a limit form of ansatz
(\ref{ansatz}), can be used to produce estimates for various characteristics of the
ground state (see, in particular, Fig. \ref{FN3} below). In fact, the TFA
will be the most relevant tool in Section V, as an analytical approximation
for trapped vortex modes, for which the use of the VA, even in its
stationary form, is too cumbersome.

The TFA cannot be applied to nonstationary solutions, hence it does not
provide direct predictions for stability of stationary modes. However, it
usually tends to produce ground states, thus predicting stable
solutions. This expectation is corroborated by results produced below.

\section{Numerical results for fundamental modes}

\begin{figure}[tb]
\includegraphics[width=0.48\columnwidth]{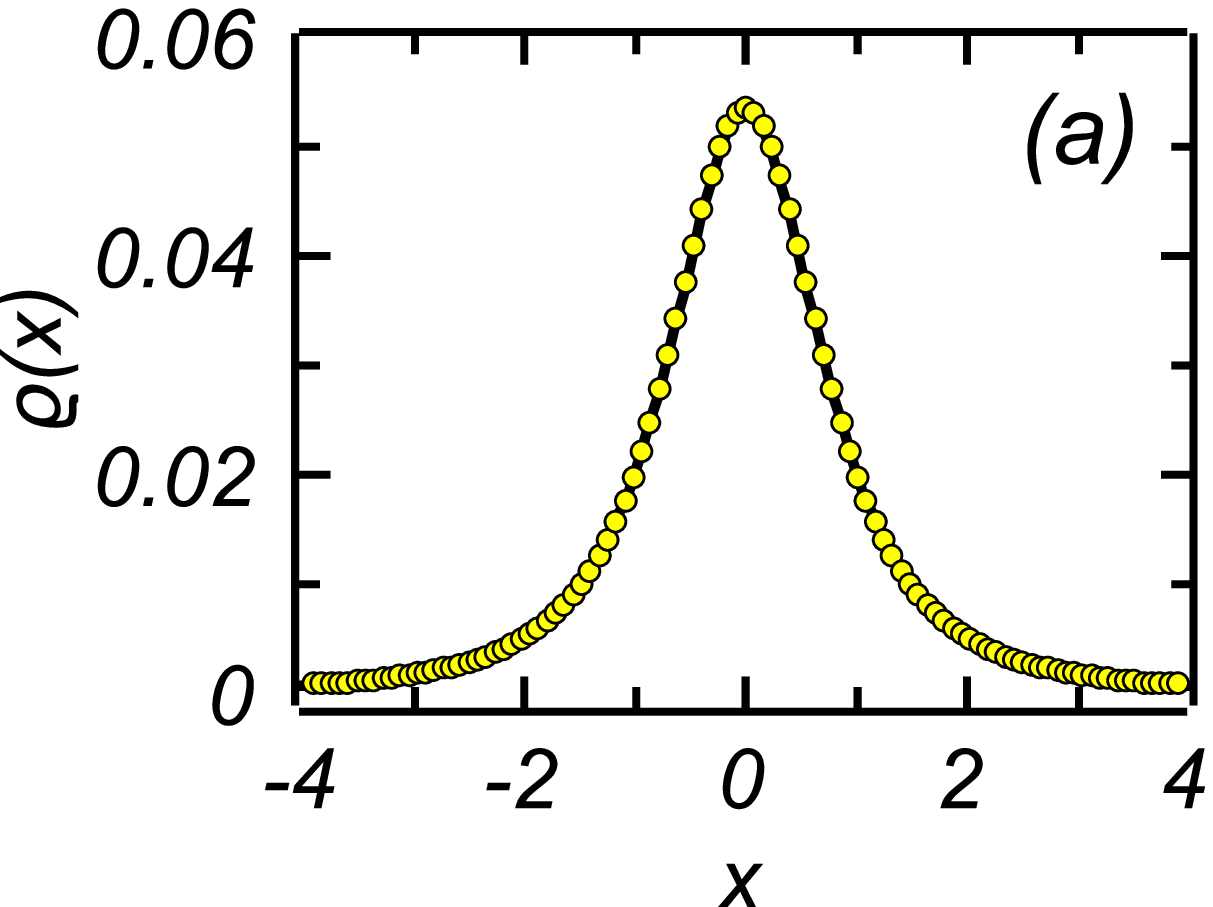} \hfil
\includegraphics[width=0.48\columnwidth]{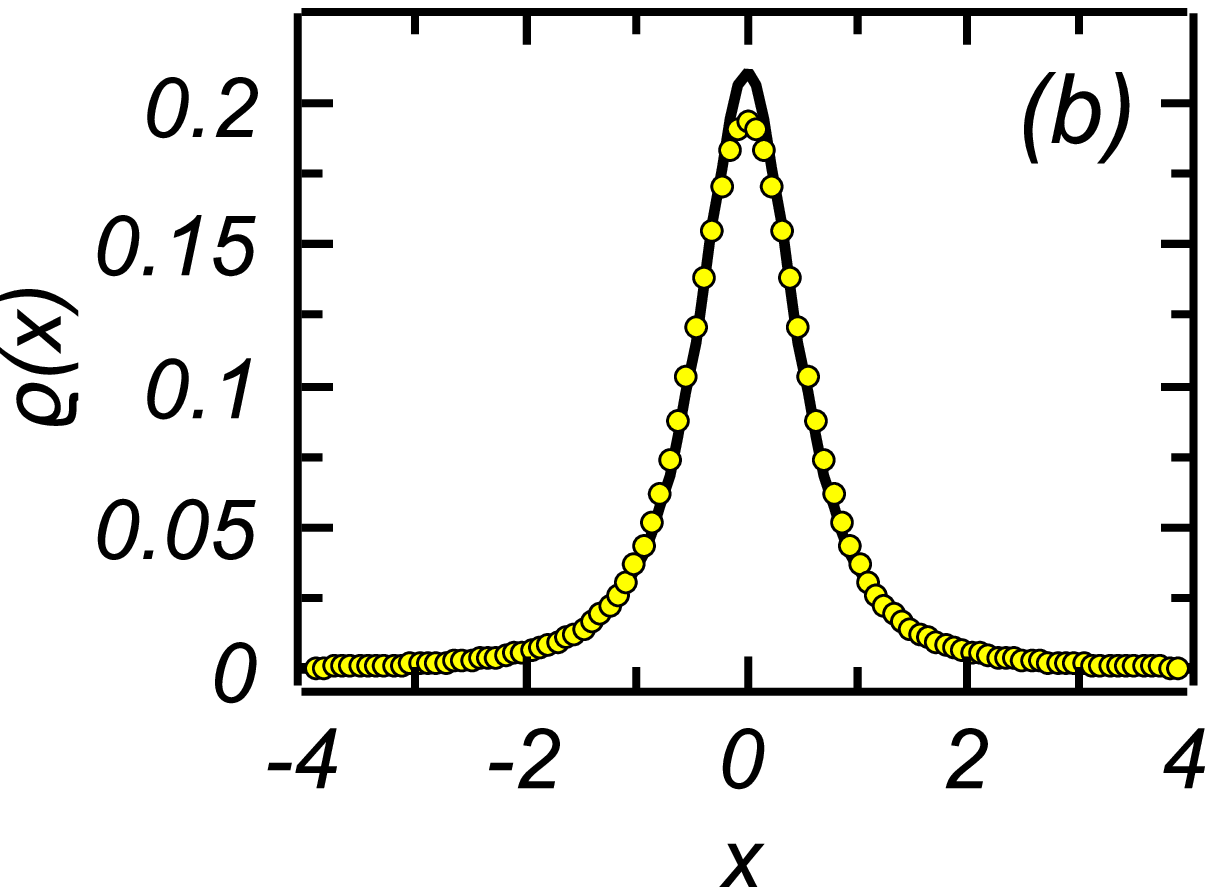} \hfil
\includegraphics[width=0.48\columnwidth]{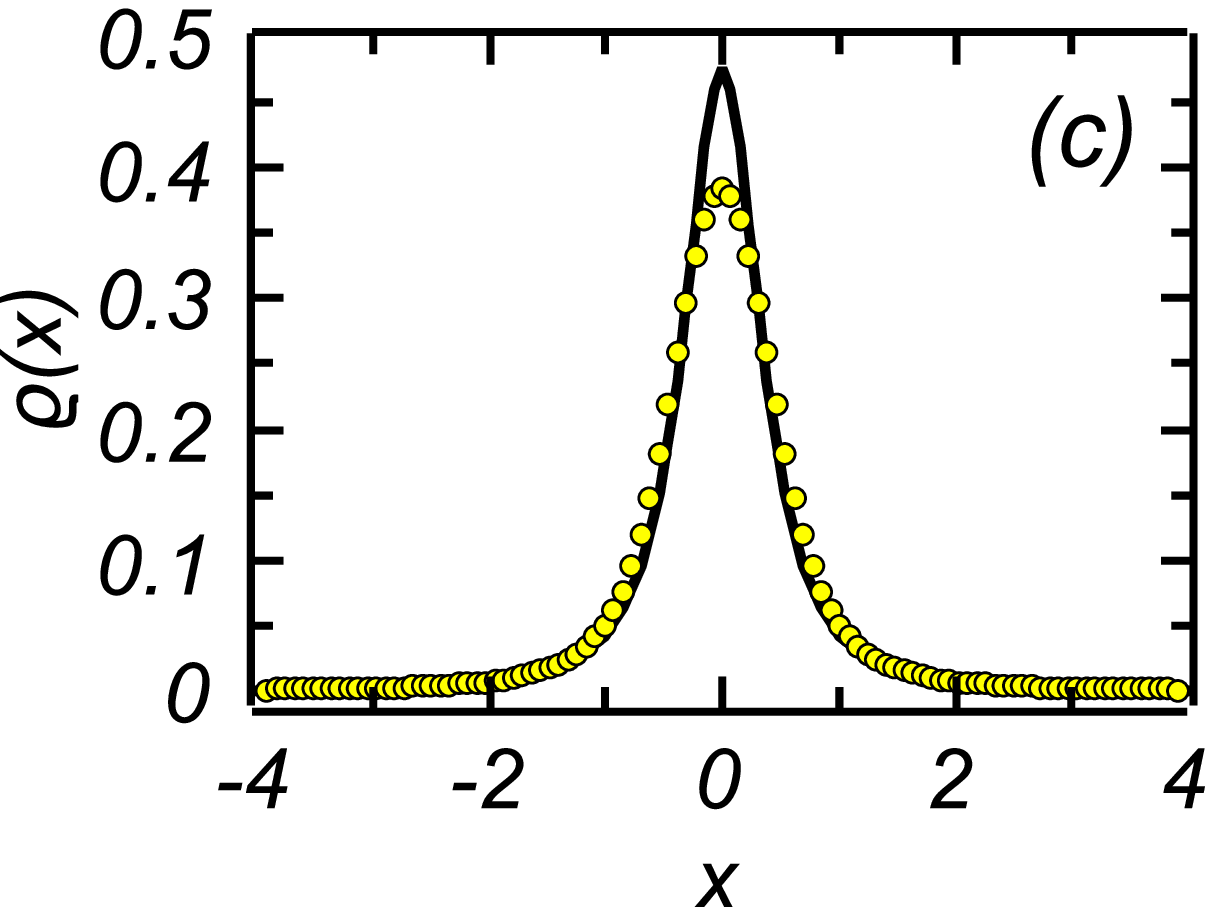}
\caption{(Color online) Profiles of fundamental trapped modes, $\protect%
\varrho (x)$, obtained via imaginary-time simulations of Eq. (\protect\ref%
{LuLe}), are shown by yellow circles. Black solid lines display counterparts
of the same profiles produced by the variational approximation based on
ansatz (\protect\ref{ansatz}). The parameters are (a) $\Delta =-1$, (b) $%
\Delta =-4$, (c) $\Delta =-10$, others fixed as $\Omega =10$, $\protect%
\gamma =1$, $E=10$, and $g=1$ (the self-defocusing nonlinearity). }
\label{FN1}
\end{figure}

\begin{figure}[tb]
\includegraphics[width=0.48\columnwidth]{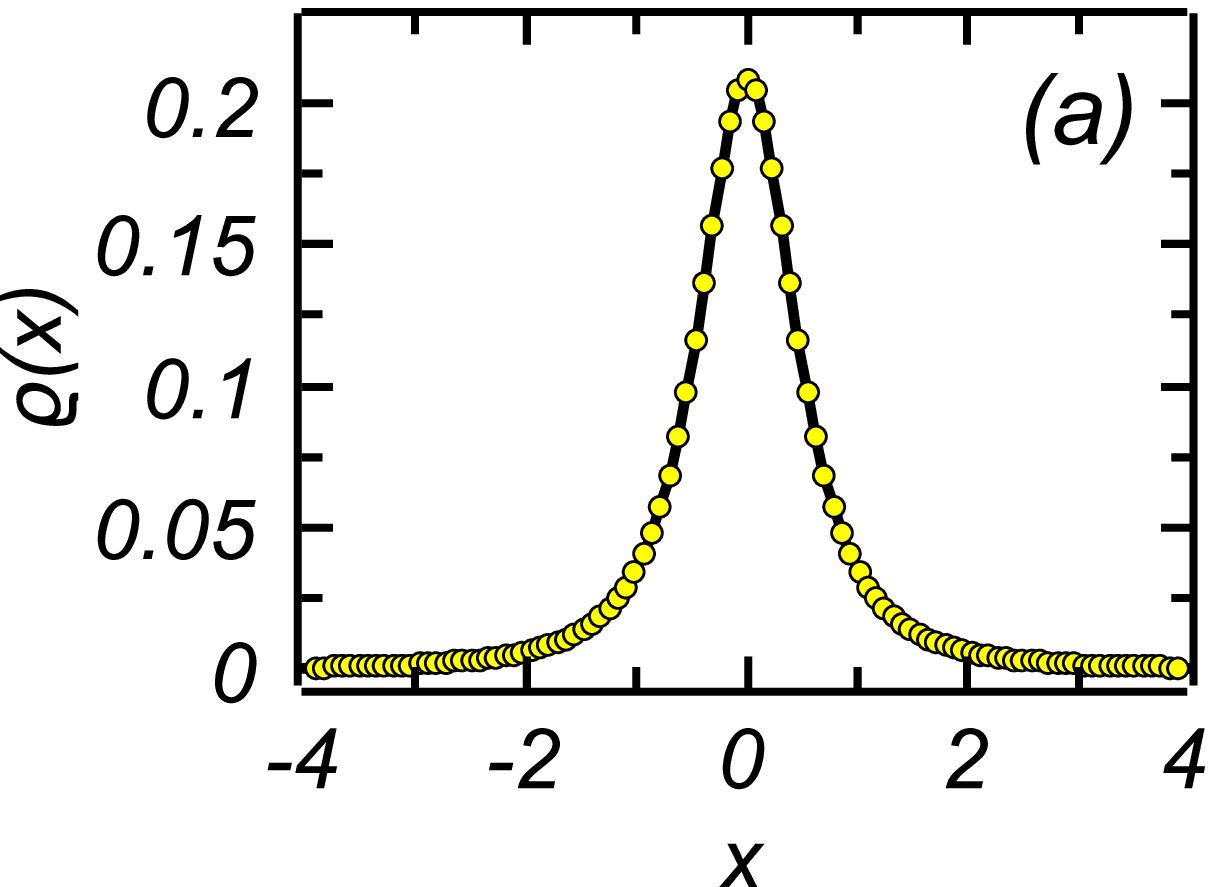} \hfil
\includegraphics[width=0.48\columnwidth]{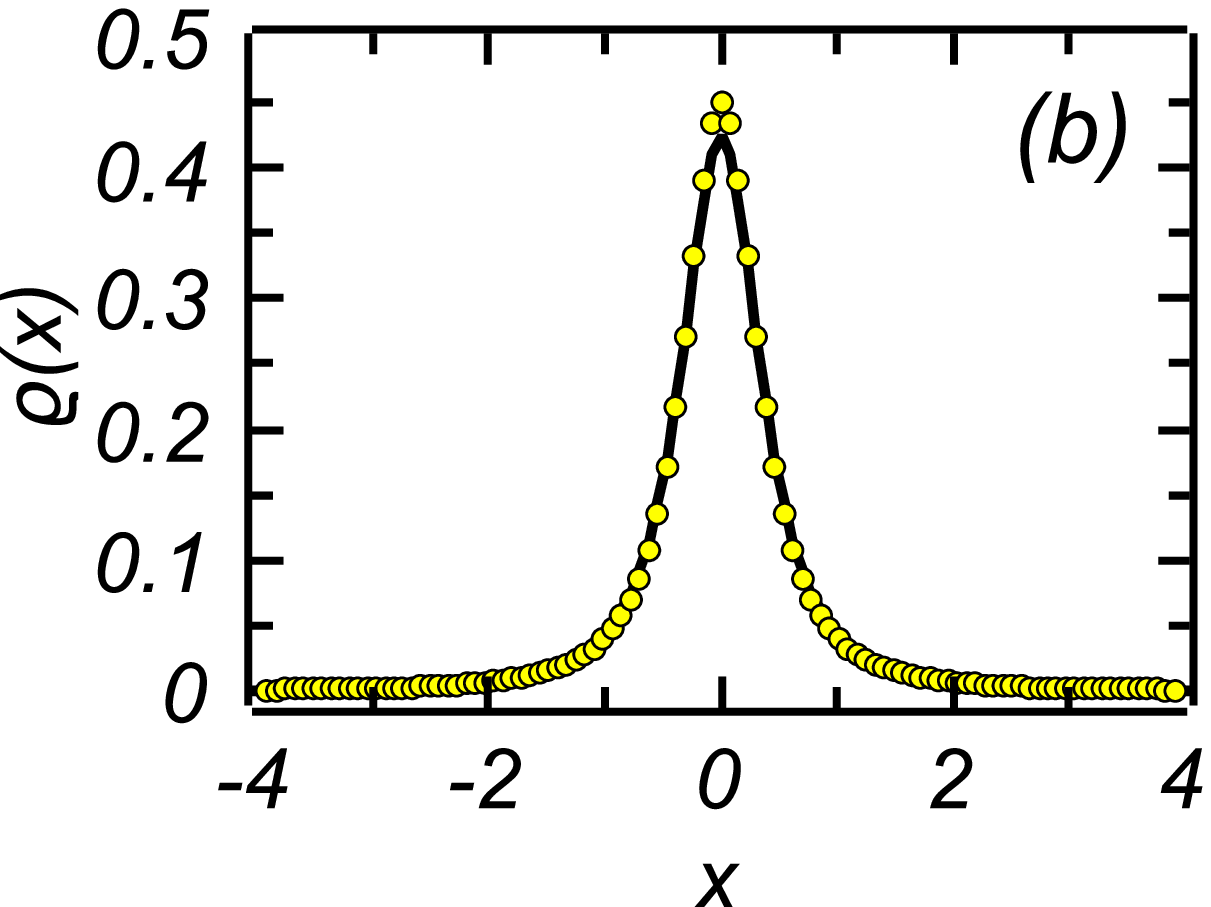} \hfil
\includegraphics[width=0.48\columnwidth]{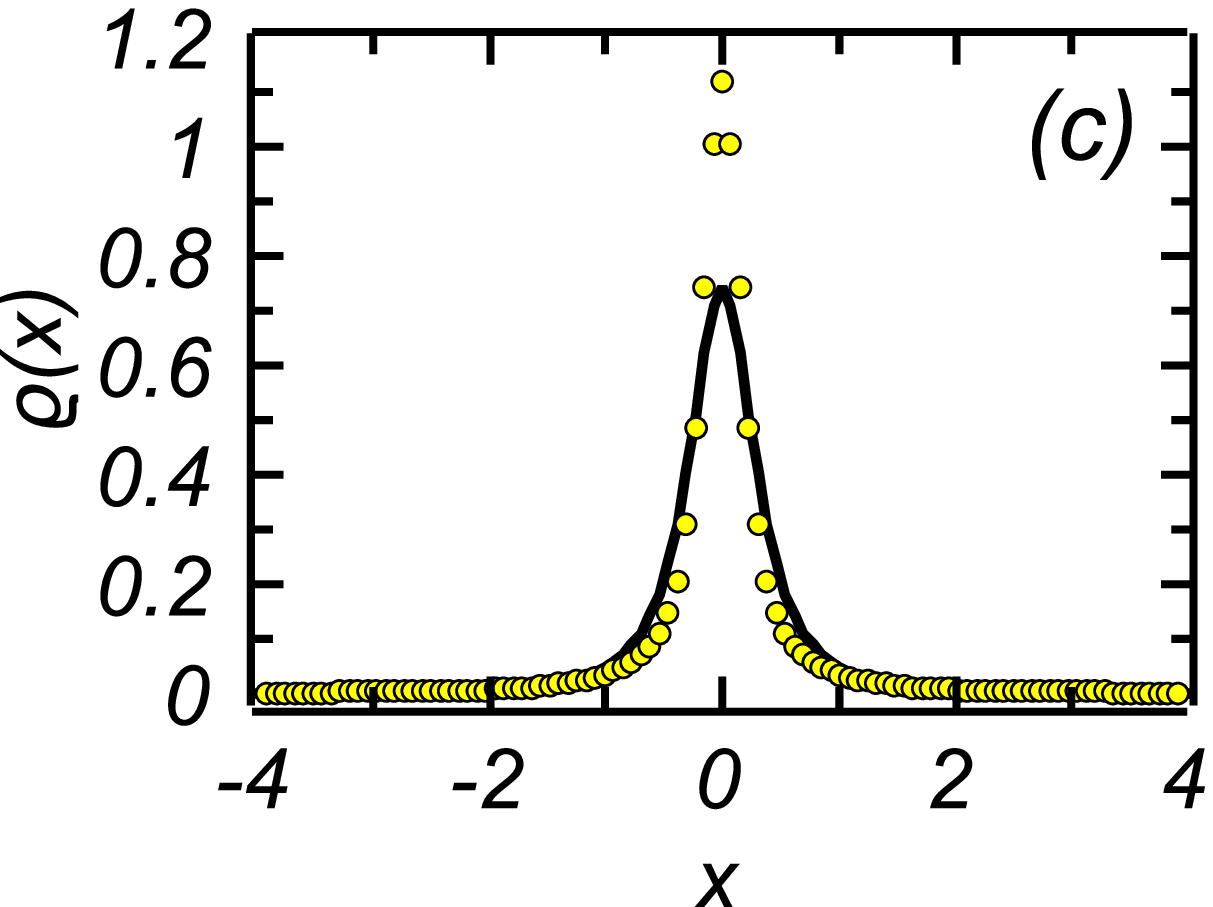}
\caption{(Color online) The same as in Fig. \protect\ref{FN1}, for
parameters (a) $\Delta =1$, (b) $\Delta =4$, (c) $\Delta =10$, with $\Omega
=10$\text{, }$\protect\gamma =1$, $E=10$, and $g=-1$ (the self-focusing
nonlinearity).}
\label{FN2}
\end{figure}

\begin{figure}[tb]
\includegraphics[width=0.48\columnwidth]{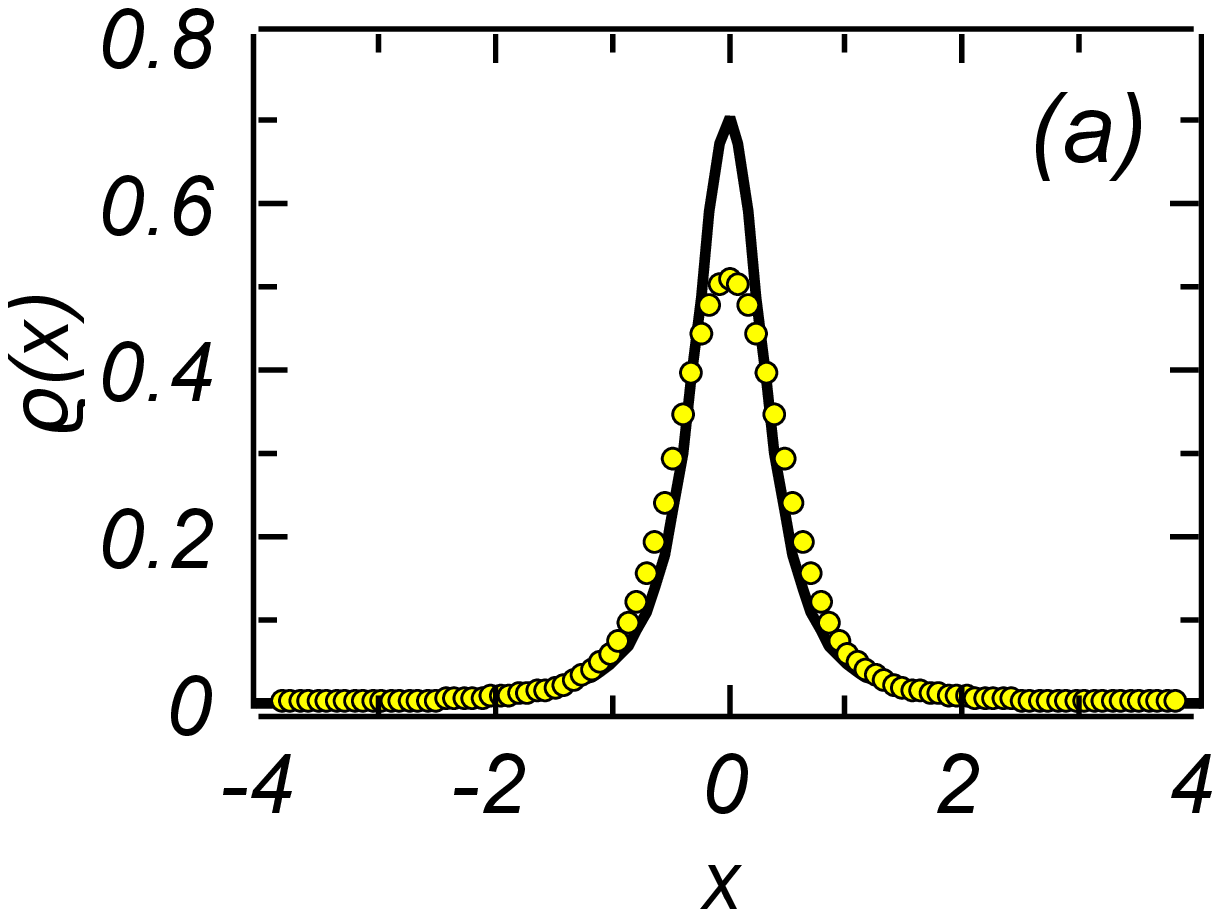} \hfil
\includegraphics[width=0.48\columnwidth]{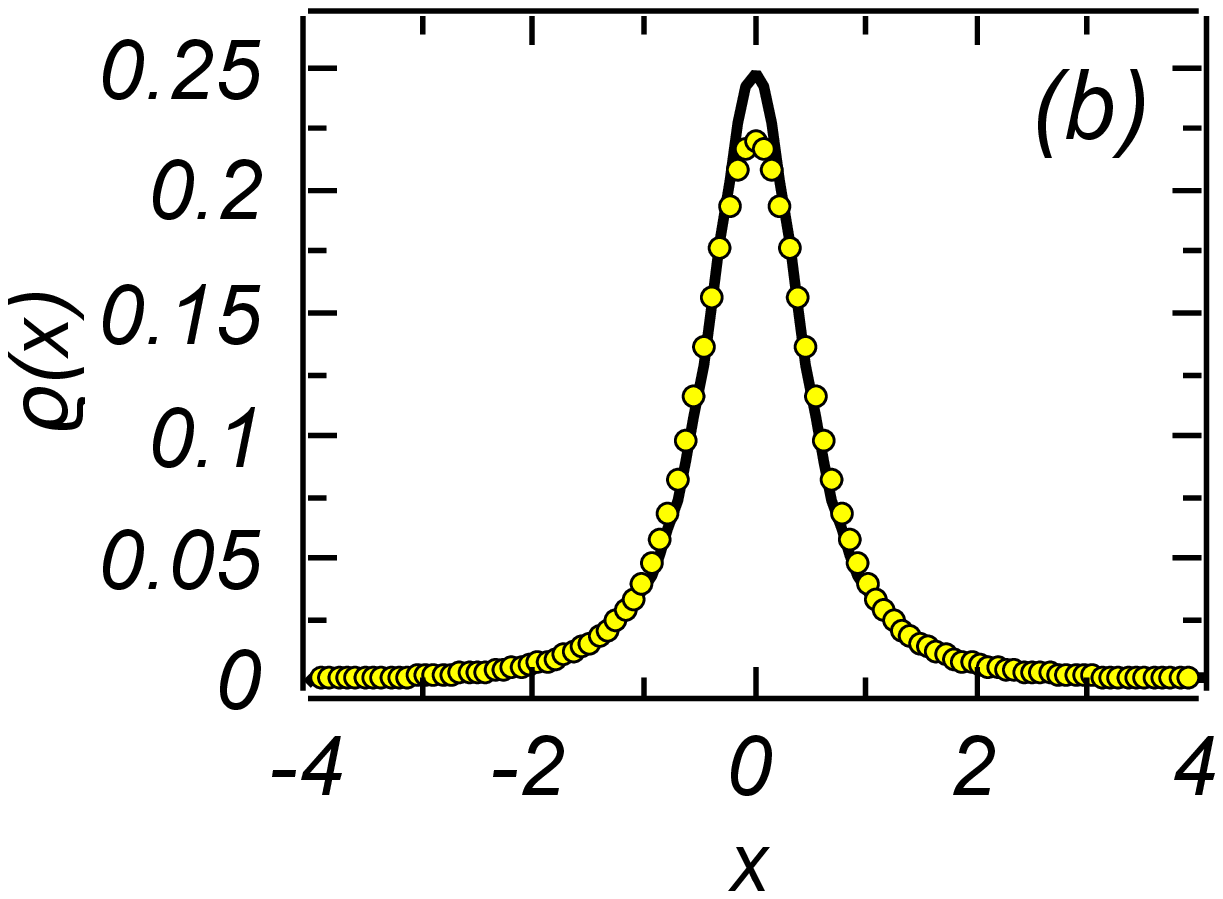}
\caption{(Color online) Following Figs. \protect\ref{FN1} and \protect\ref{FN2}, chains of
yellow circles depict profiles of fundamental trapped modes, $\protect%
\varrho (x)$, obtained via imaginary-time simulations of Eq. (\protect\ref%
{LuLe}), for the self-defocusing nonlinearity, $g=1$, and large positive
values of the mismatch: $\Delta =10$ in (a) and $\Delta =20$ in (b). Black
solid lines display the same profiles, as produced by the simplest version
of the Thomas-Fermi approximation, given by Eq. (\protect\ref{rhoTFA}).
Other parameters are $\Omega =10$, $\protect\gamma =1$, and $E=10$.}
\label{FN3}
\end{figure}

\subsection{Stationary trapped modes}

To obtain accurate results, and verify the validity of the VA predictions
which are presented in the previous section, we here report results obtained
as numerical solutions of Eq. (\ref{LuLe}). First, we aim to find
ground-state localized states by means of imaginary-time propagation. In the
framework of this method, one numerically integrates Eq. (\ref{LuLe}),
replacing $t$ by $-it$ and normalizing the solution at each step of the time
integration to maintain a fixed total power \cite{IT}. For testing stability
of stationary states, Eq. (\ref{LuLe}) was then simulated in real time, by
means of the fourth-order split-step method implemented in a \emph{GNU
Octave } program \cite{Eaton_Octave} (for a details concerning the method
and its implementations in MATLAB, see Ref. \cite{Yang_10}).

In Fig. \ref{FN1} we show 1D integrated intensity profiles $\varrho (x)$,
defined as
\begin{equation}
\varrho (x)\equiv \int_{-\infty }^{+\infty }|\phi (x,y)|^{2}dy,  \label{rho}
\end{equation}%
and obtained from the imaginary-time solution of Eq. (\ref{LuLe}), along with their
analytical counterparts produced by the VA based on Eq. (\ref{ansatz}), for
three different values of detuning $\Delta $, \textit{viz}., (a) $\Delta =-1$%
, (b) $\Delta =-4$, and (c) $\Delta =-10$, for $g=1$ (the self-defocusing
nonlinearity) and $\Omega =10$ and $E=10$. We used Eqs. (\ref{ff-simple})
and (\ref{th-simple}) to produce values of the parameters $f$ and $\chi $ in
ansatz (\ref{ansatz}), which was then used as the initial guess in direct
numerical simulations.

In Fig. \ref{FN2} we display results similar to those shown in Fig. \ref{FN1}%
, but for the self-focusing nonlinearity ($g=-1$) and three different
(positive) values of $\Delta $, \textit{viz.}, (a) $\Delta =1$, (b) $\Delta
=4$, and (c) $\Delta =10$, with fixed $\Omega =10$ and $E=10$. In both cases
of $g=\pm 1$, the VA profiles show good match to the numerical ones,
although the accuracy slightly deteriorates with the increase of $|\Delta |$%
.
\begin{figure}[tb]
\centering
\includegraphics[width=0.48\columnwidth]{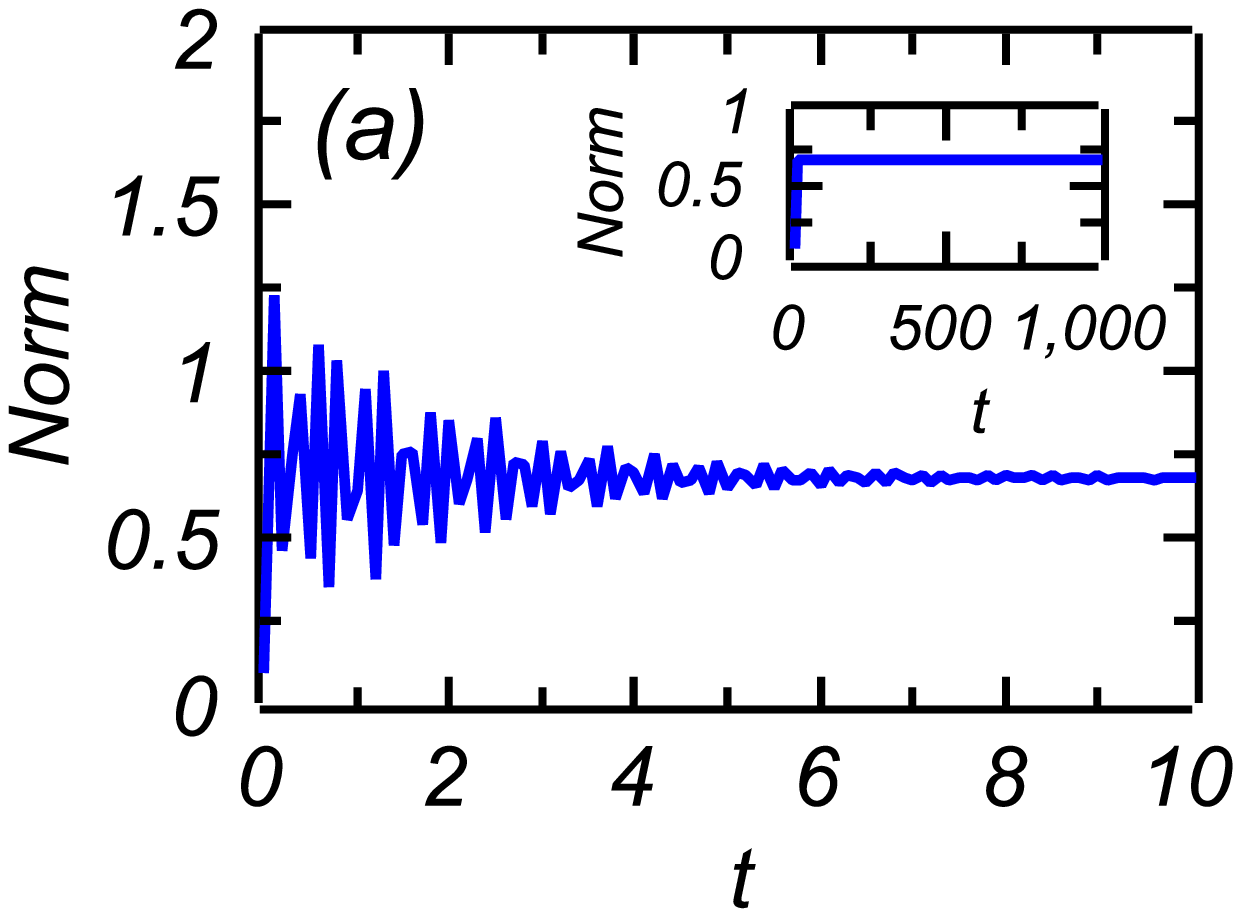} \hfil %
\includegraphics[width=0.48\columnwidth]{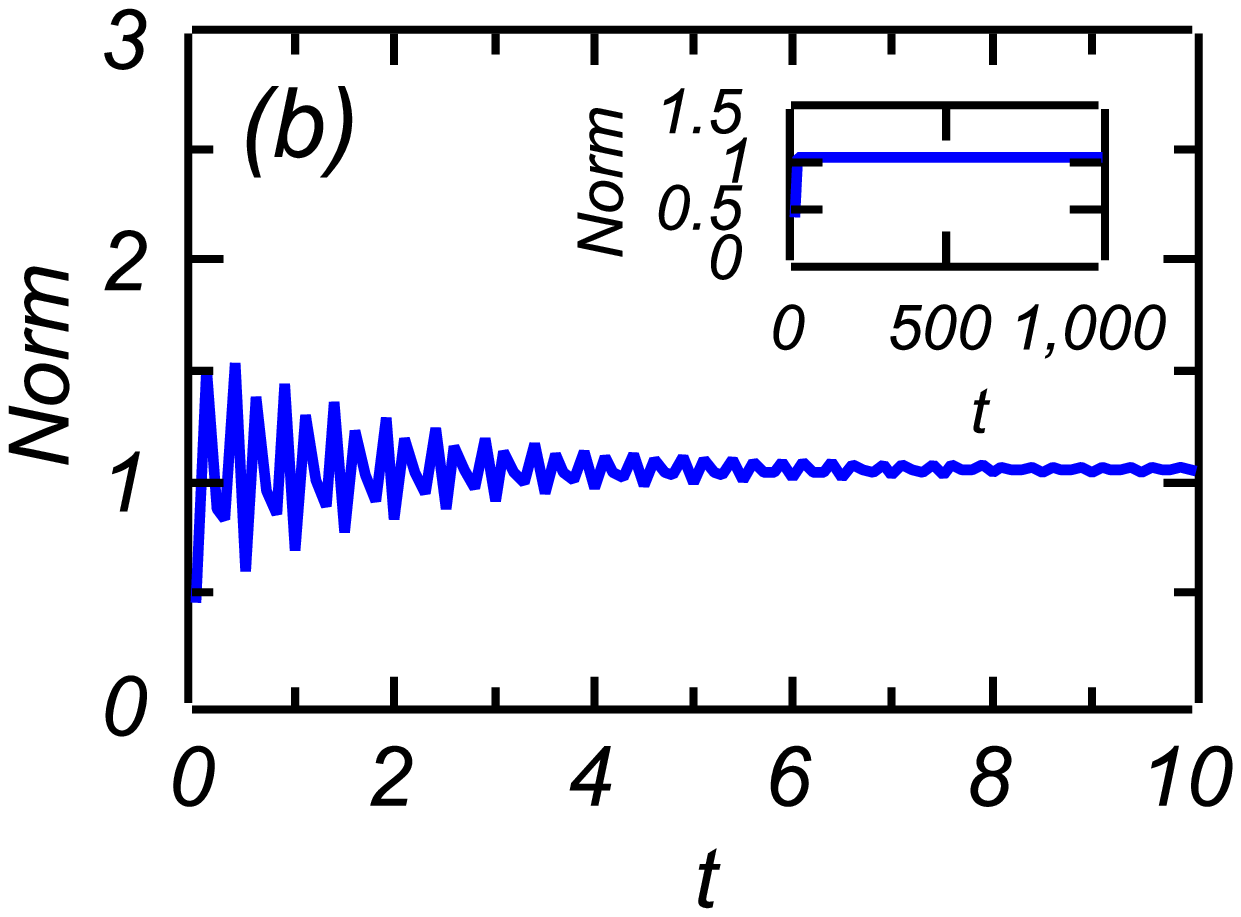} %
\includegraphics[width=0.48\columnwidth]{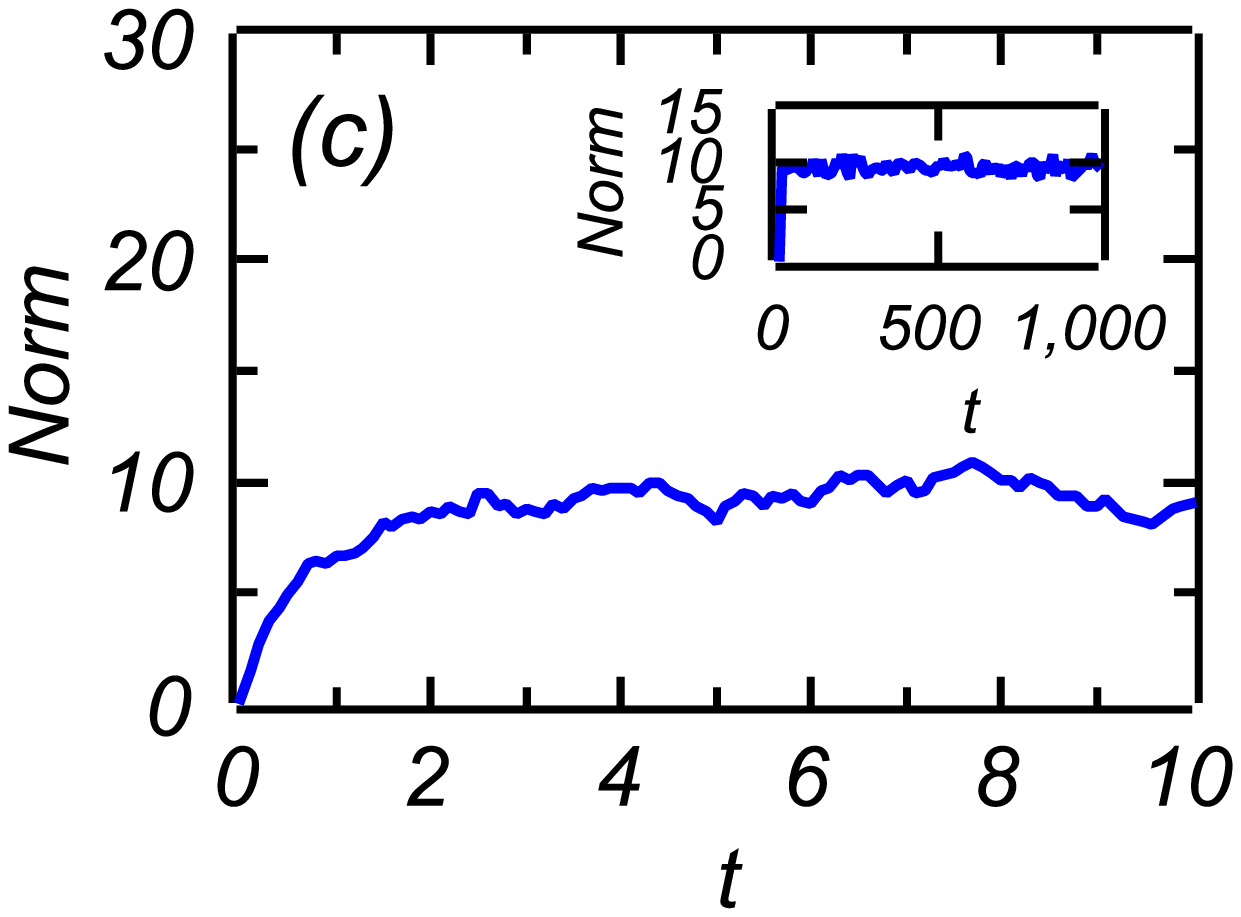} \hfil
\includegraphics[width=0.48\columnwidth]{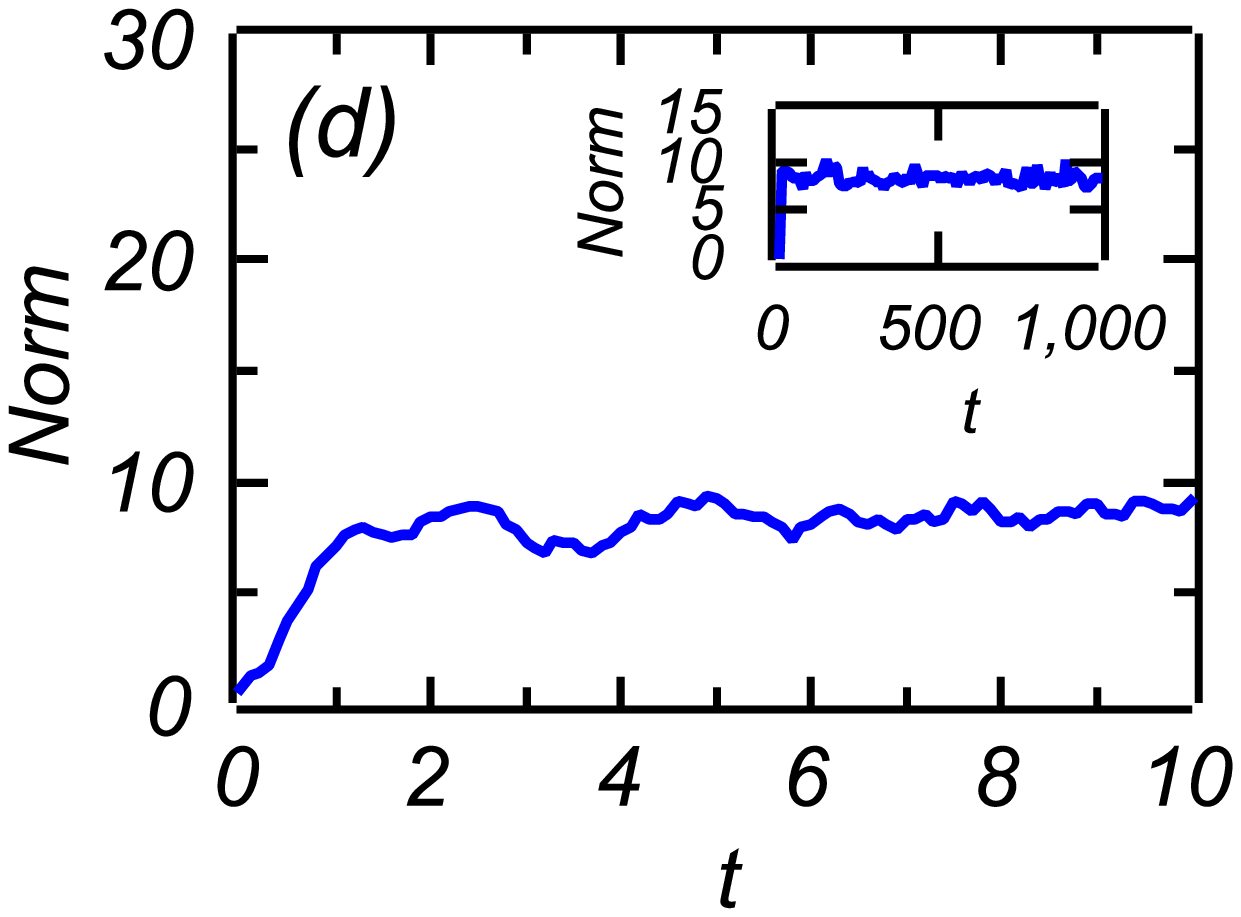}
\caption{(Color online) The evolution of the norm [total power (\protect\ref%
{Norm})] of the solution starting from ansatz (\protect\ref{ansatz})
perturbed by $5\%$ random noise, as produced by real-time simulations of Eq.
(\protect\ref{LuLe}). Here the results are presented for the defocusing
nonlinearity, i.e., $g=1$, with (a) $\Delta =-1$ and (b) $\Delta =-10$, and
for the focusing nonlinearity, i.e., $g=-1$, with (c) $\Delta =1$ and (d) $%
\Delta =10$. Other parameters are $E=10$, $\Omega =10$, and $\protect\gamma %
=1$. The evolution of the norm at large times is shown in the insets.}
\label{fig_norm}
\end{figure}

\begin{figure}[tb]
\centering
\includegraphics[width=0.45\columnwidth]{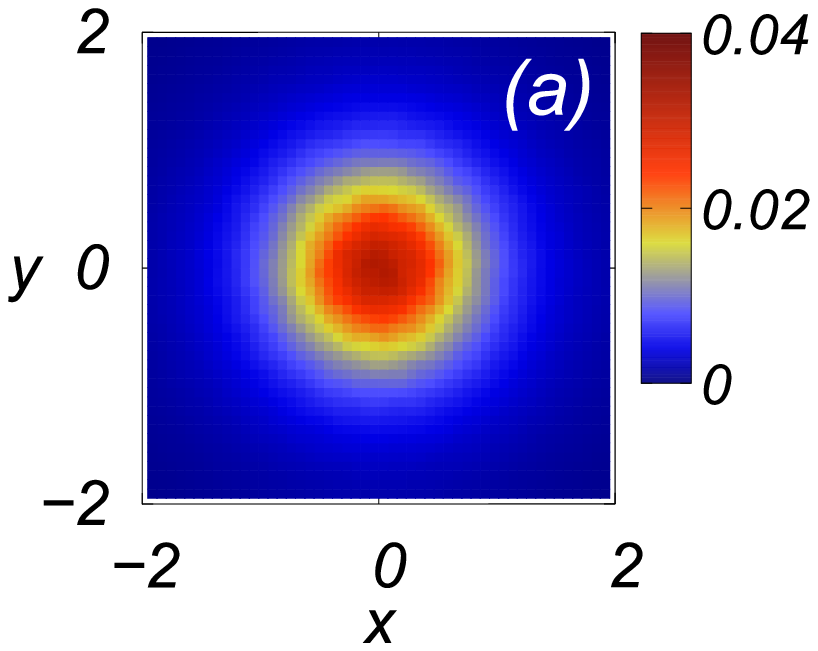} \hfil
\includegraphics[width=0.45\columnwidth]{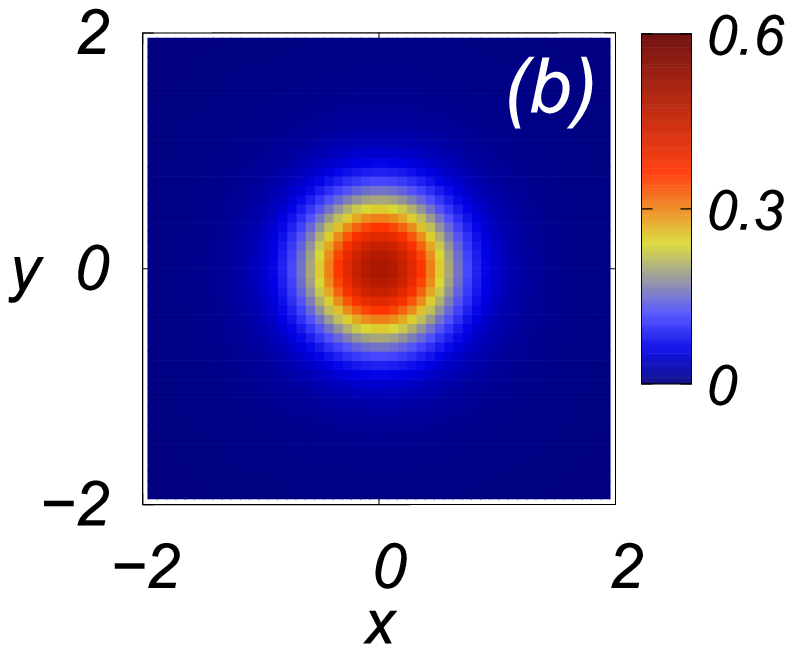} \hfil
\includegraphics[width=0.45\columnwidth]{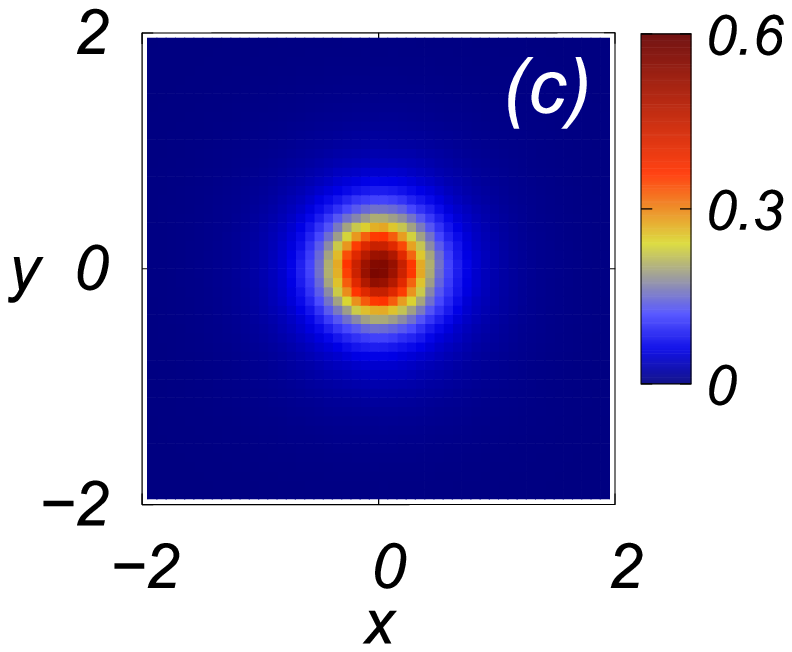} \hfil
\includegraphics[width=0.45\columnwidth]{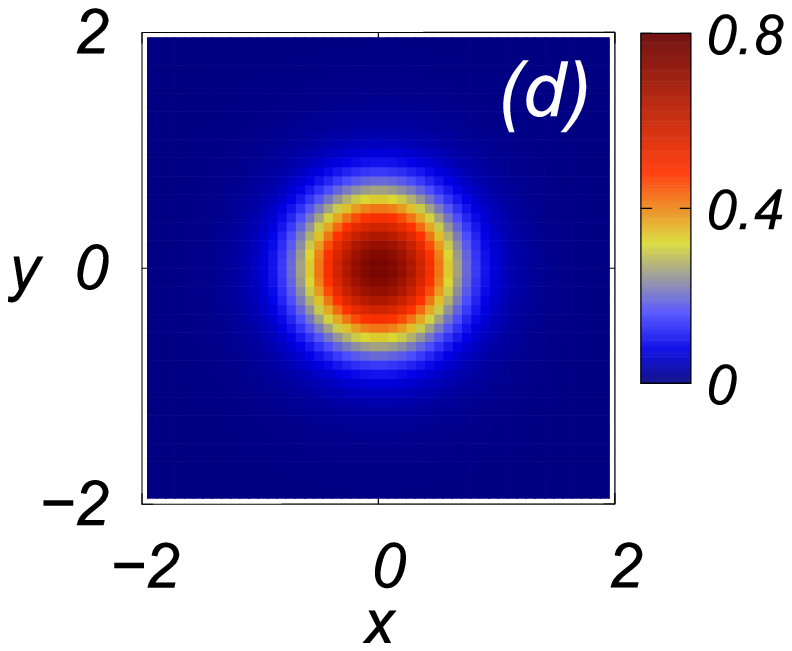}
\caption{(Color online) Density profile $|\protect\phi |^{2}$ obtained via
direct numerical simulations of Eq. (\protect\ref{LuLe}) in the case of the
self-defocusing nonlinearity ($g=1$). Inputs, represented by ansatz (\protect
\ref{ansatz}) with the addition of $5\%$ random noise, are displayed in
panels (a) for $\Delta =-1$ and (c) for $\Delta =-10$. The corresponding
profiles produced by the simulations at $t=1000$ are shown in (b) and (d),
respectively. Other parameters are the same as in Fig. \protect\ref{fig_norm}%
.}
\label{perf_def}
\end{figure}

Note that the results displayed in Fig. \ref{FN2}, for the situations to
which the TFA does not apply, because the nonlinearity is self-focusing in
this case, demonstrate \ the growth of the maximum value, $\varrho (x=0)$,
with the increase of mismatch $\Delta $. In the case of self-defocusing it
is natural to expect decay of $\varrho _{\mathrm{TFA}}(x=0)$ with the
increase of $\Delta $. As shown in Fig. \ref{FN3}, this expectation is
confirmed by the numerical results and the TFA alike. In particular, for the
integrated intensity profile defined by Eq. (\ref{rho}), the simplest
version of the TFA, produced by Eq. (\ref{simple}), easily gives%
\begin{equation}
\varrho _{\mathrm{TFA}}(x)=\frac{2\pi E^{2}}{\Omega ^{4}}\left( \frac{%
2\Delta }{\Omega ^{2}}+x^{2}\right) ^{-3/2}.  \label{rhoTFA}
\end{equation}%
Figure \ref{FN3} also corroborates that the TFA, even in its simplest form,
becomes quite accurate for sufficiently large values of $\Delta >0$.

\subsection{Stability of the stationary modes}

\begin{figure}[tb]
\centering
\includegraphics[width=0.45\columnwidth]{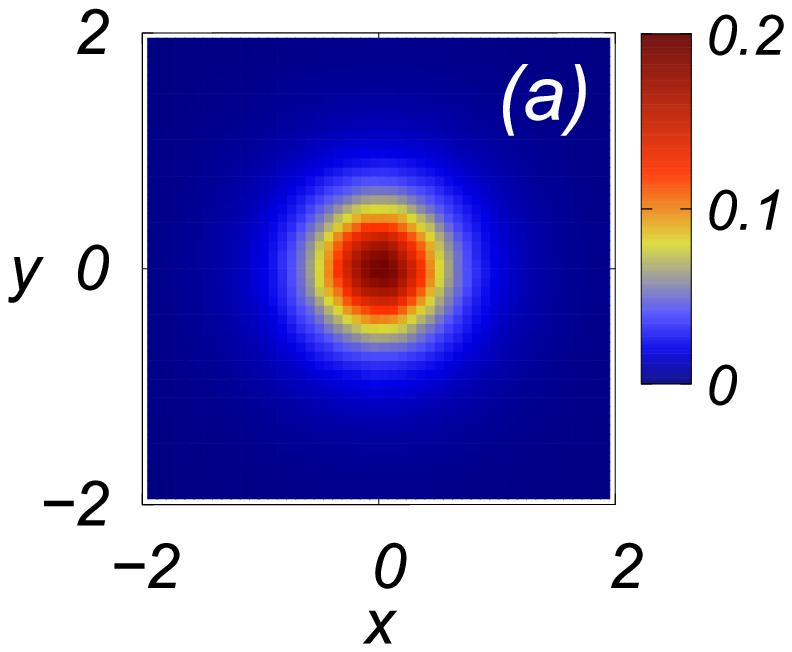} \hfil
\includegraphics[width=0.45\columnwidth]{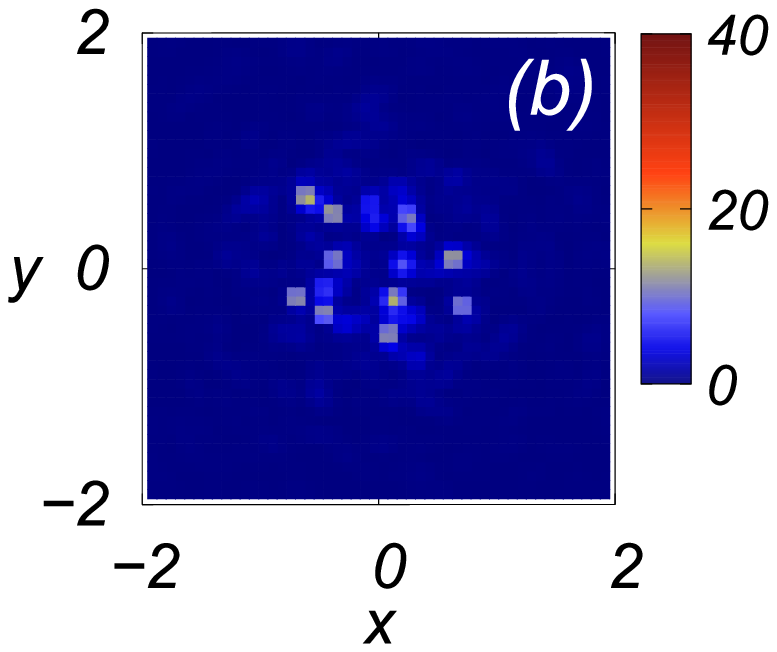} \hfil
\includegraphics[width=0.45\columnwidth]{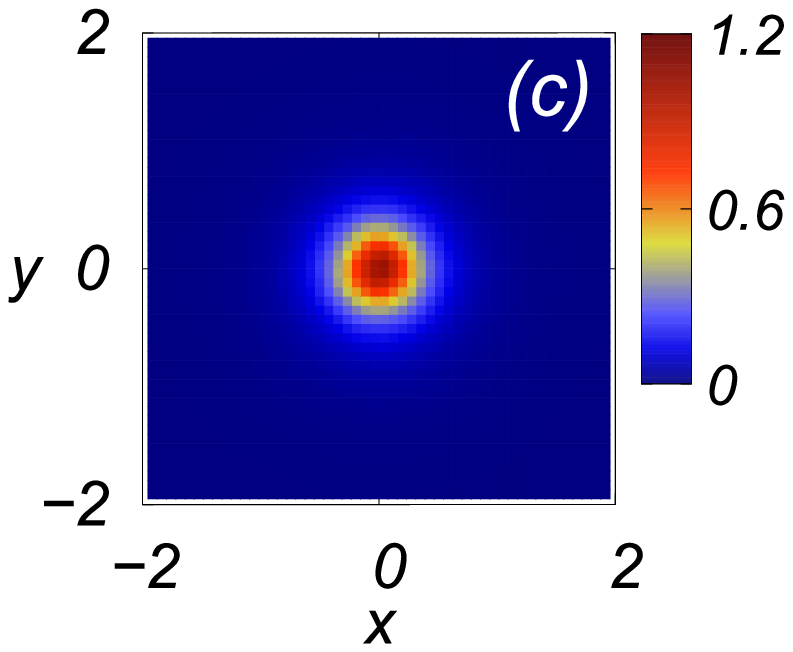} \hfil
\includegraphics[width=0.45\columnwidth]{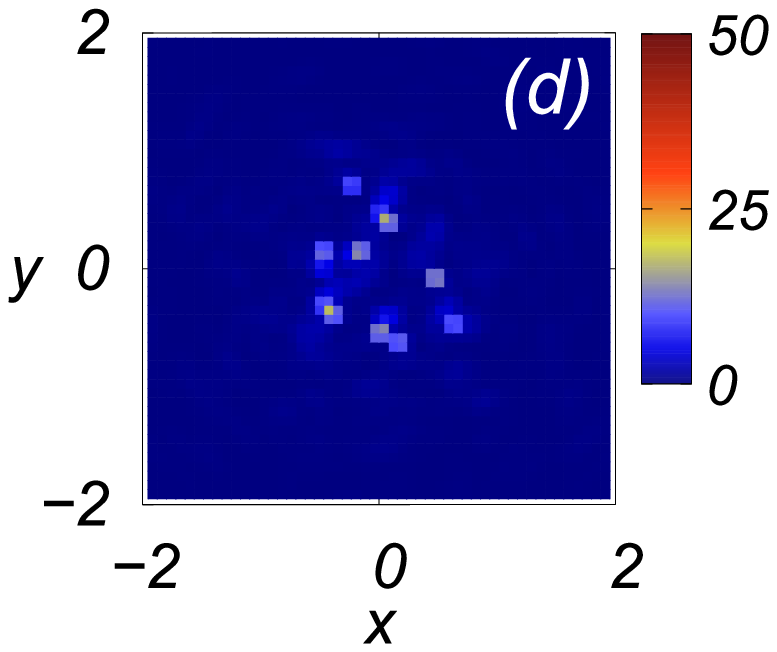}
\caption{(Color online) The same as in Fig. \protect\ref{perf_def}, but in
the case of the self-focusing nonlinearity ($g=-1$). Panels (a,b) and (c,d)
are drawn for $\Delta =1$ and $\Delta =10$, respectively. Here, the profiles
shown in (b) and (d) are outputs of the simulations obtained at $t=10$.}
\label{perf_foc}
\end{figure}

The stability of the trapped configurations predicted by ansatz (\ref{ansatz}%
) was tested in real-time simulations of Eq. (\ref{LuLe}), adding $5\%$
random noise to the input. We display the results, showing the evolution of
the solution's norm (total power) in the case of the defocusing nonlinearity
($g=1$), for $\Delta =-1$ and $-10$, in Figs. \ref{fig_norm}(a) and (b),
respectively. The insets show the asymptotic behavior at large times. In the
case of the defocusing nonlinearity, the solution quickly relaxes to a
numerically exact stationary form, and remains \emph{completely stable} at $%
t>10$ (in fact, real-time simulations always quickly converge to stable
solutions at all values of the parameters). However, in the case of the
self-focusing with $\Omega =10$, the solutions are unstable, suffering rapid
fragmentation, as seen in Fig. \ref{perf_foc}. This behavior is also
exemplified in results shown in Figs. \ref{fig_norm}(c) and \ref{fig_norm}%
(d) for the temporal evolution of the solution's total power in the case of
the self-focusing nonlinearity ($g=-1$), for $\Delta =1$ and for $\Delta =10$%
, respectively. The instability of the fundamental modes in this case is a
natural manifestation of the modulational instability in the LL equation
\cite{MI}. Note that the large size of local amplitudes in small spots,
which is attained in the course of the development of the instability
observed in Fig. \ref{perf_foc}, implies the trend to the onset of the 2D
collapse driven by the self-focusing cubic nonlinearity \cite{collapse}.

\begin{figure}[tb]
\centering
\includegraphics[width=0.6\columnwidth]{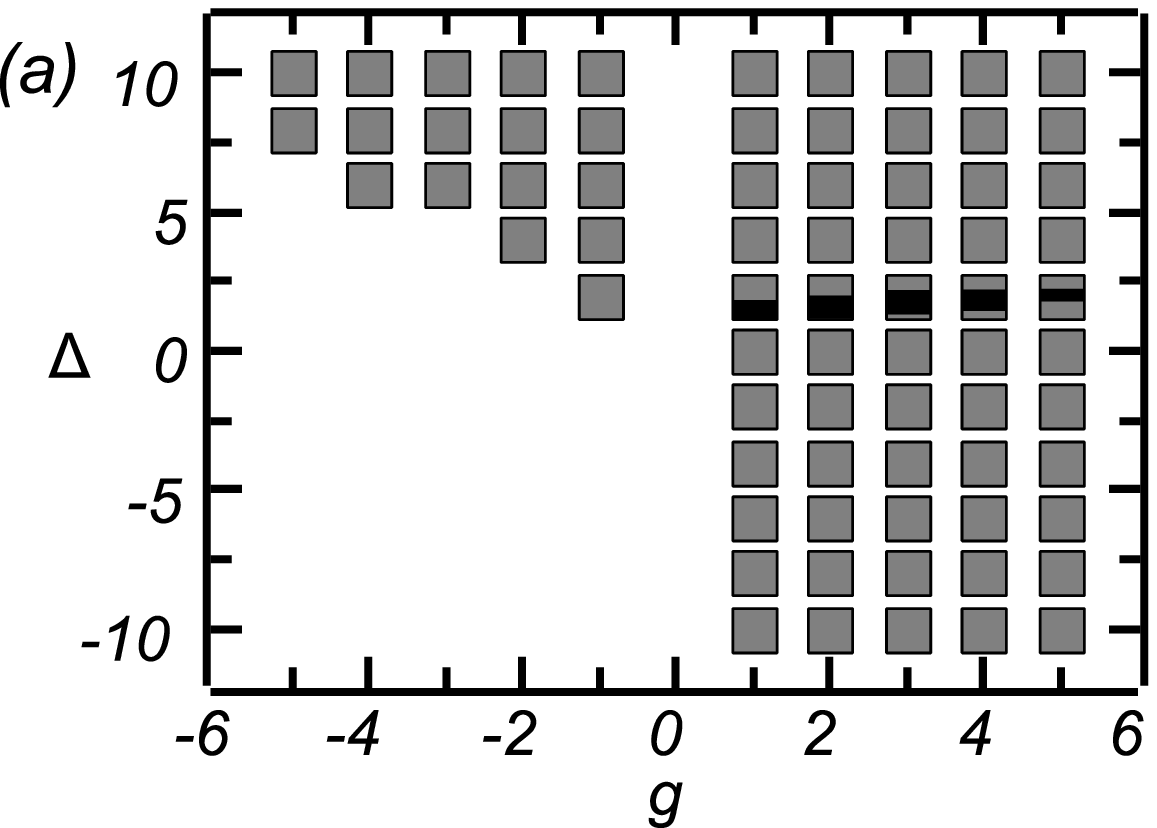} \\
\includegraphics[width=0.6\columnwidth]{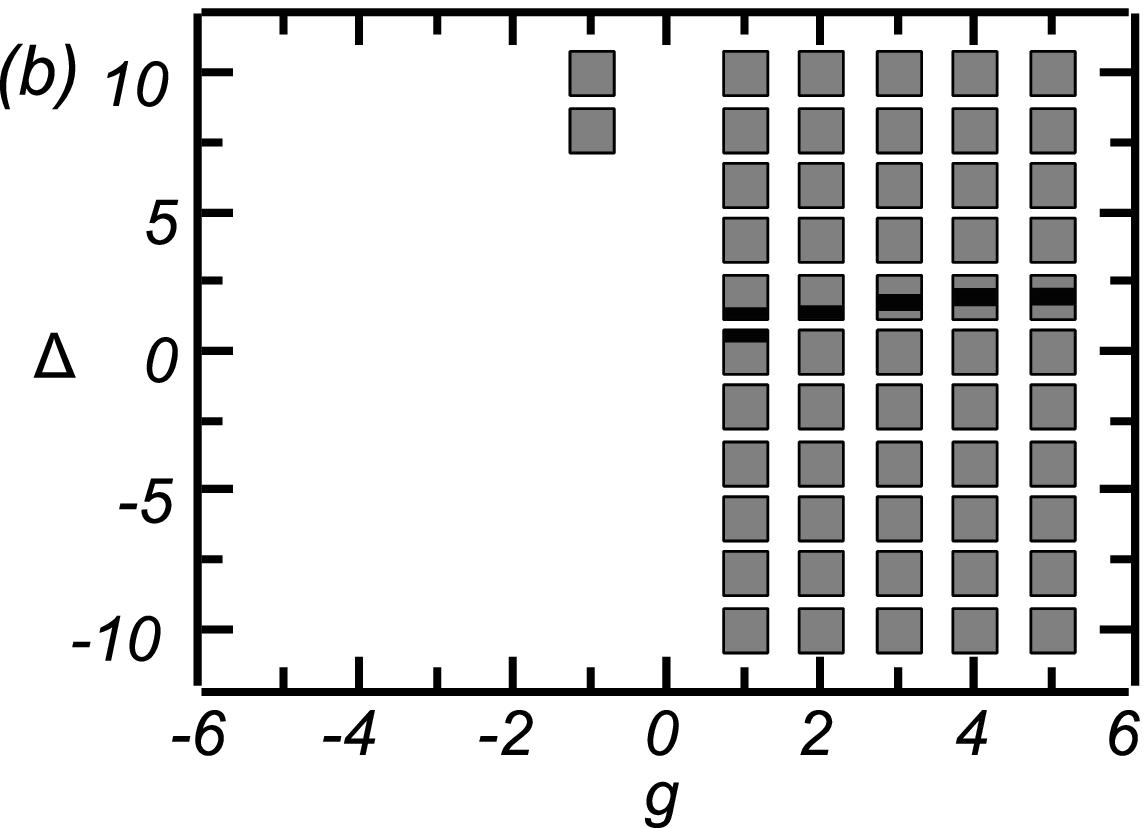} \\
\includegraphics[width=0.6\columnwidth]{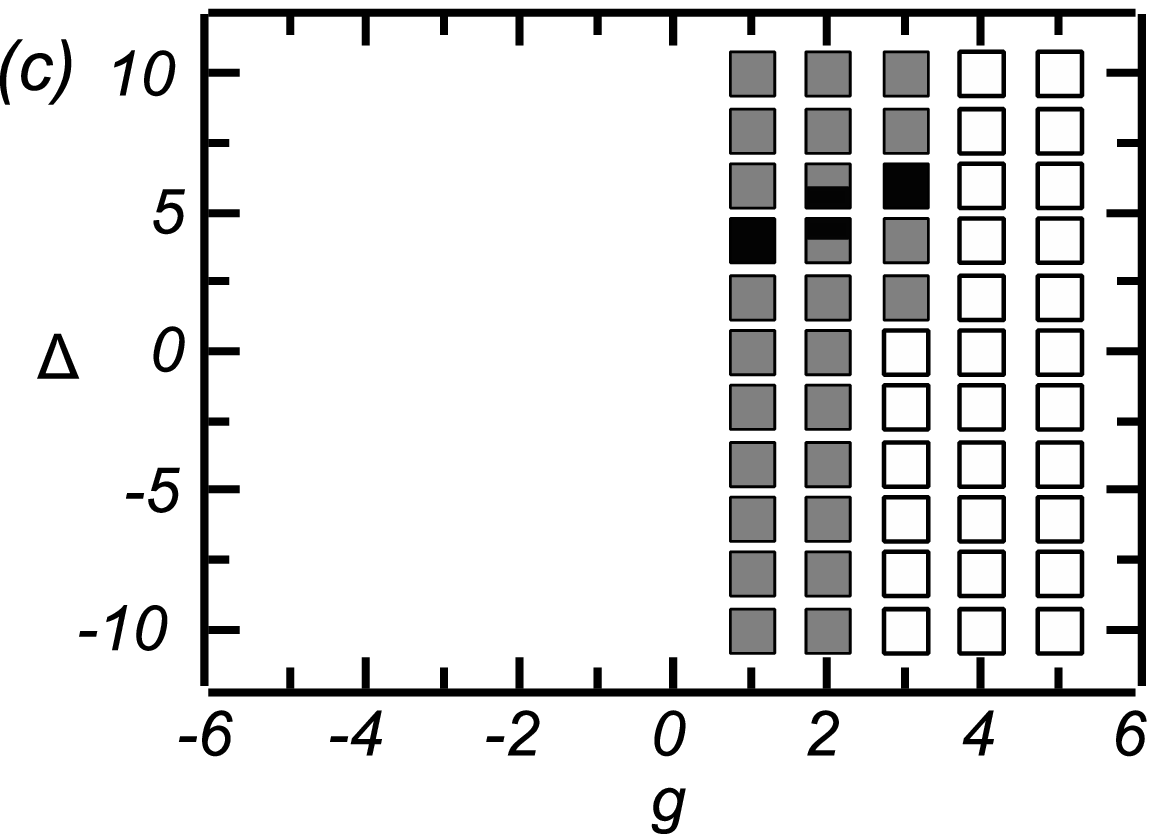}
\caption{The existence area of stable modes obtained by means of real-time
simulation of Eq. (\protect\ref{LuLe}). To generate the area, we used the
input in the form of ansatz (\protect\ref{ansatz}) with parameters predicted
by the VA, adding random noise at the $5\%$ amplitude level. We here consider
(a) $\Omega =2$, (b) $\Omega =4$, and (c) $\Omega =10$. In all the cases,
the norm (total power) of the solution undergoes variations. For parameter
values corresponding to the gray boxes it stabilizes after a short
relaxation period. In the white boxes, the norm keeps oscillating, while the
solution maintains the localized profile, avoiding onset of instability. The
simulations in the region not covered by boxes feature instability
scenarios: in the case of $g\leq 0$ the solution suffers fragmentation, like
in Figs. \protect\ref{perf_foc}(b) and \protect\ref{perf_foc}(d)), while in
the case of self-defocusing the solution is subject to fragmentation due to
the strong nonlinearity [e.g., at $g>5$ in (c)]. In the black region, the
output states are very close to the input. Other parameters are $E=10$, and $%
\protect\gamma =1$. The data for the linear system, corresponding to $g=0$,
are not included, as in the linear system all the stationary solutions are
obviously stable.}
\label{NER}
\end{figure}

\begin{figure}[tb]
\includegraphics[width=0.48\columnwidth]{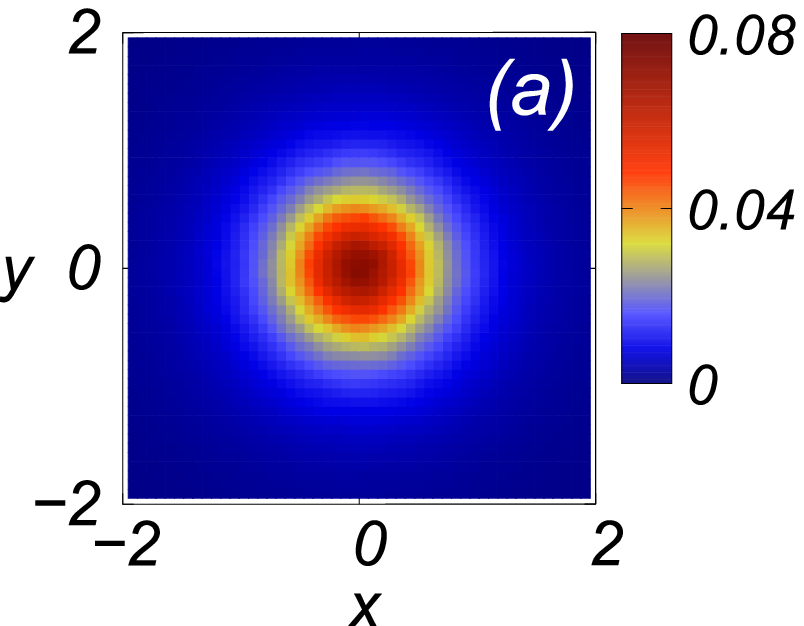} \hfil
\includegraphics[width=0.48\columnwidth]{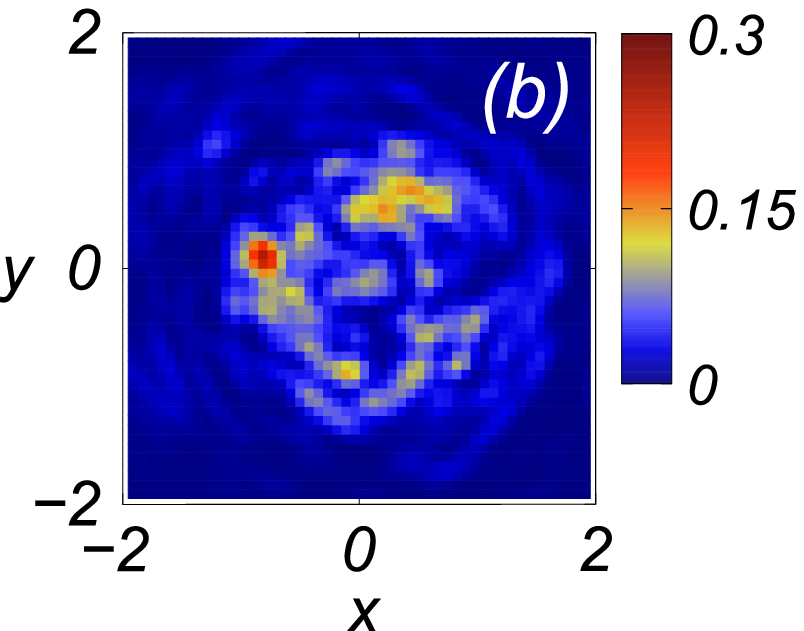} %
\includegraphics[width=0.48\columnwidth]{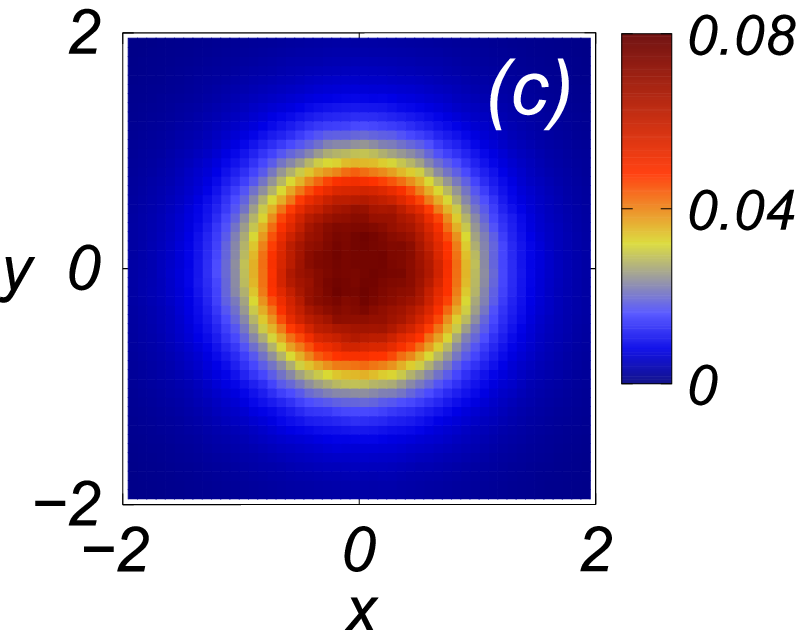} \hfil
\includegraphics[width=0.48\columnwidth]{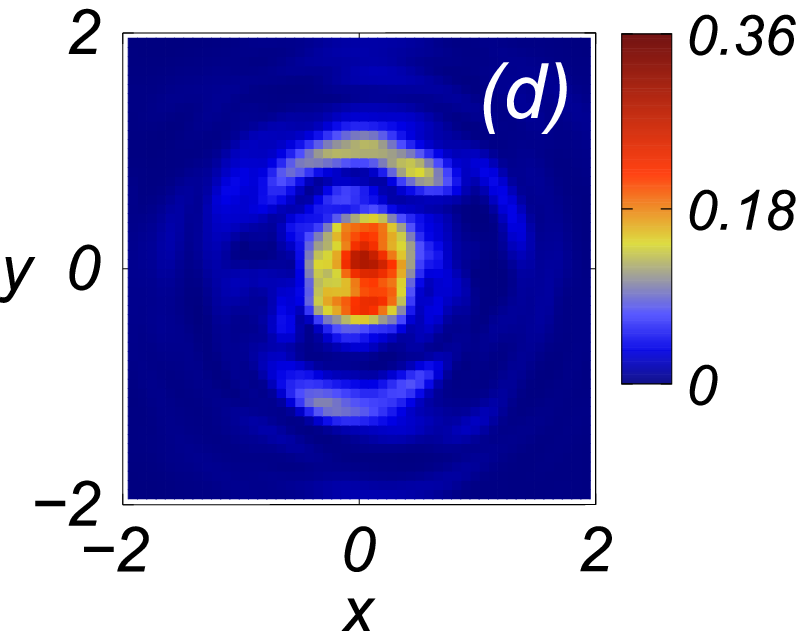}
\caption{(Color online) The same as in Fig. \protect\ref{perf_foc}, but for $%
g=10$. Panels (a,b) and (c,d) pertain to $\Delta =-10$ and $\Delta =10$,
respectively. The unstable output profiles are displayed at $t=10$.}
\label{F9}
\end{figure}

In Figs. \ref{perf_def} and \ref{perf_foc} we display the time evolution of
density profiles $|\phi |^{2}$ produced by the simulations of Eq. (\ref{LuLe}%
) with the self-defocusing and focusing nonlinearity, respectively. The
input profiles are again taken as per the VA ansatz (\ref{ansatz}) with the
addition of $5\%$ random noise. In Figs. \ref{perf_def}(a) and \ref{perf_def}%
(c) we show the perturbed input profiles in the case of self-defocusing, for
$\Delta =-1$ and $\Delta =-10$, respectively, while the corresponding
profiles at $t=1000$ are displayed in Figs. \ref{perf_def}(b) and \ref%
{perf_def}(d). Note that the agreement between the variational and numerical
profiles tends to deteriorate with the increase of $|\Delta |$ [the same
trend as observed in Fig. \ref{F5}(c)].

Further, in Figs. \ref{perf_foc}(a) and \ref{perf_foc}(c) we display the
perturbed input profiles in the case of the self-focusing nonlinearity for $%
\Delta =1$ and $\Delta =10$, respectively, with the corresponding profiles
at $t=1000$ displayed in Figs. \ref{perf_foc}(b) and \ref{perf_foc}(d).
These results clearly confirm the instability of the perturbed solutions, as
suggested by the evolution of the total power depicted in Figs. \ref%
{fig_norm}(c) and \ref{fig_norm}(d). Strong instability is observed for all
values of $g<0$, which corresponds to the self-focusing.

The findings for the existence and stability of the localized pixels are
summarized by diagrams displayed in Fig. \ref{NER}. To produce them, we
analyzed the temporal evolution of total power (\ref{Norm}), parallel to
monitoring the spatial profile of each solution at large times ($t=100$ and $%
t=1000$). In Fig. \ref{NER}, we address three different values of strength $%
\Omega $ of the trapping potential: (a) $\Omega =2$, (b) $\Omega =4$, and
(c) $\Omega =10$. The stability area is represented by gray and white boxes,
which correspond, respectively, to robust static outputs and those which
feature small residual oscillations, while the parameter area not covered by 
boxes corresponds to unstable solutions. This includes the area of $g\leq
0$ (self-focusing), where the modes suffer strong instability observed in
Figs. \ref{perf_foc}(b) and \ref{perf_foc}(d) at $\Omega =10$, but may be
stable at $\Omega =2$ and $4$ [in the latter case, the stability domain for $%
g>0$ is very small, as seen in Fig. \ref{NER}(b)]. On the other hand, at $%
g>5$ and $\Omega =10$, the solution undergoes fragmentation under the action
of the strong self-defocusing nonlinearity, for all values of $-10\leq
\Delta \leq +10$. An example of that is displayed in Fig. \ref{F9} for $g=10$
and two extreme values of the mismatch, $\Delta =-10$ and $+10$.

In the stability area, black spots highlight values of the parameters at
which the output profiles of the static solutions, observed at $t=1000$, are
very close to the respective input profiles, i.e., the VA provides very
accurate predictions. Generally, the shape of the stability area in the form
of the vertical stripe, observed in Fig. \ref{NER}(c), roughly follows the
vertical direction of the dotted black line in Fig. \ref{ER}, which pertains
to the same value of $\Omega =10$. On the other hand, the expansion of the
stability area in the horizontal direction for $\Omega =2$ and $\Omega =4$,
which is observed in Figs. \ref{NER}(a,b), qualitatively complies with the
strong change of the curves in Fig. \ref{ER} for the same values of $\Omega $%
. Looking at Fig. \ref{NER}, one can also conclude that large positive
values of $\Delta $ help to additionally expand the stability region.

We stress that the results shown in Fig. \ref{NER} are extremely robust:
real-time simulations lead to them, even starting with zero input. The input
provided by the VA ansatz (\ref{ansatz}) is used above to explore the
accuracy of the VA, which is relevant, as similar approximations can be
applied to similar models, incorporating the pump, linear loss, and Kerr
nonlinearity (self-defocusing or focusing).

\section{Vortex solitons}

\subsection{Analytical considerations: the Thomas-Fermi approximation}

\begin{figure}[tb]
\includegraphics[width=0.48\columnwidth]{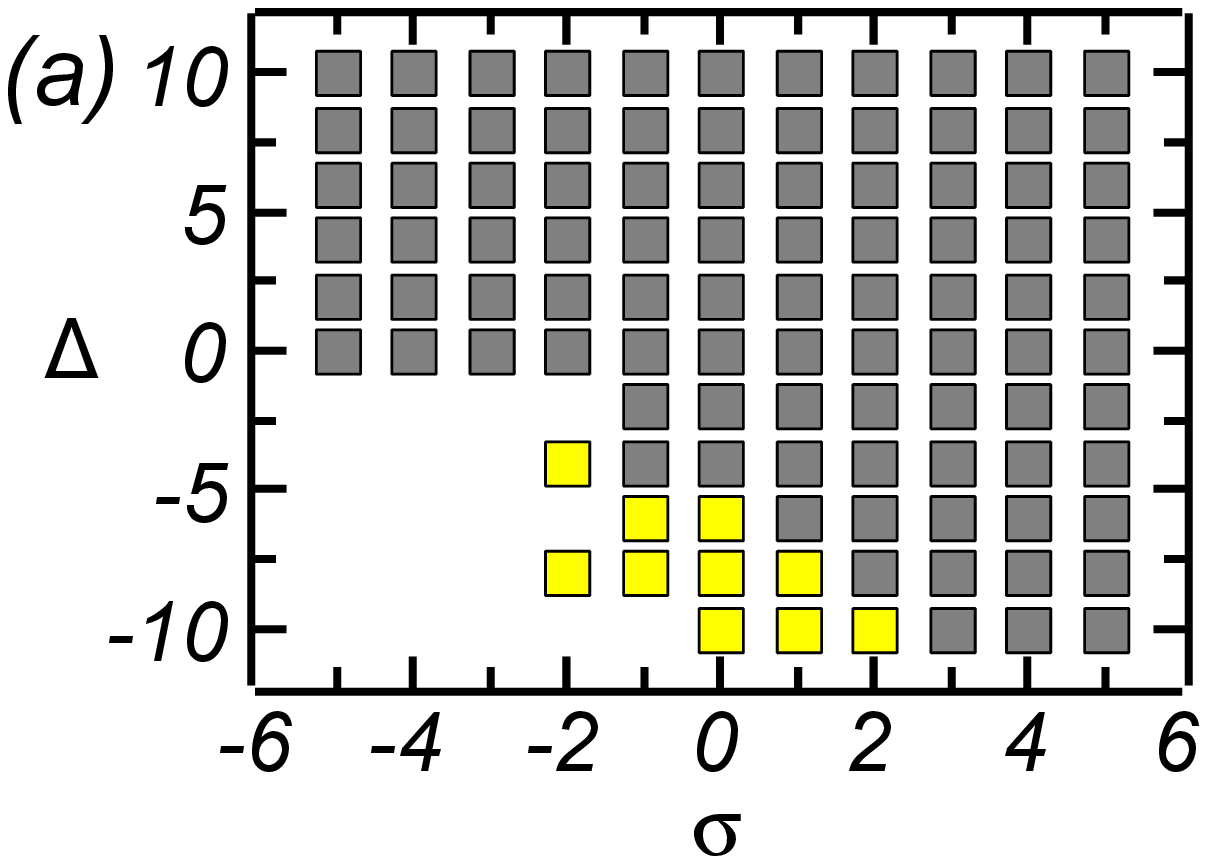} %
\includegraphics[width=0.48\columnwidth]{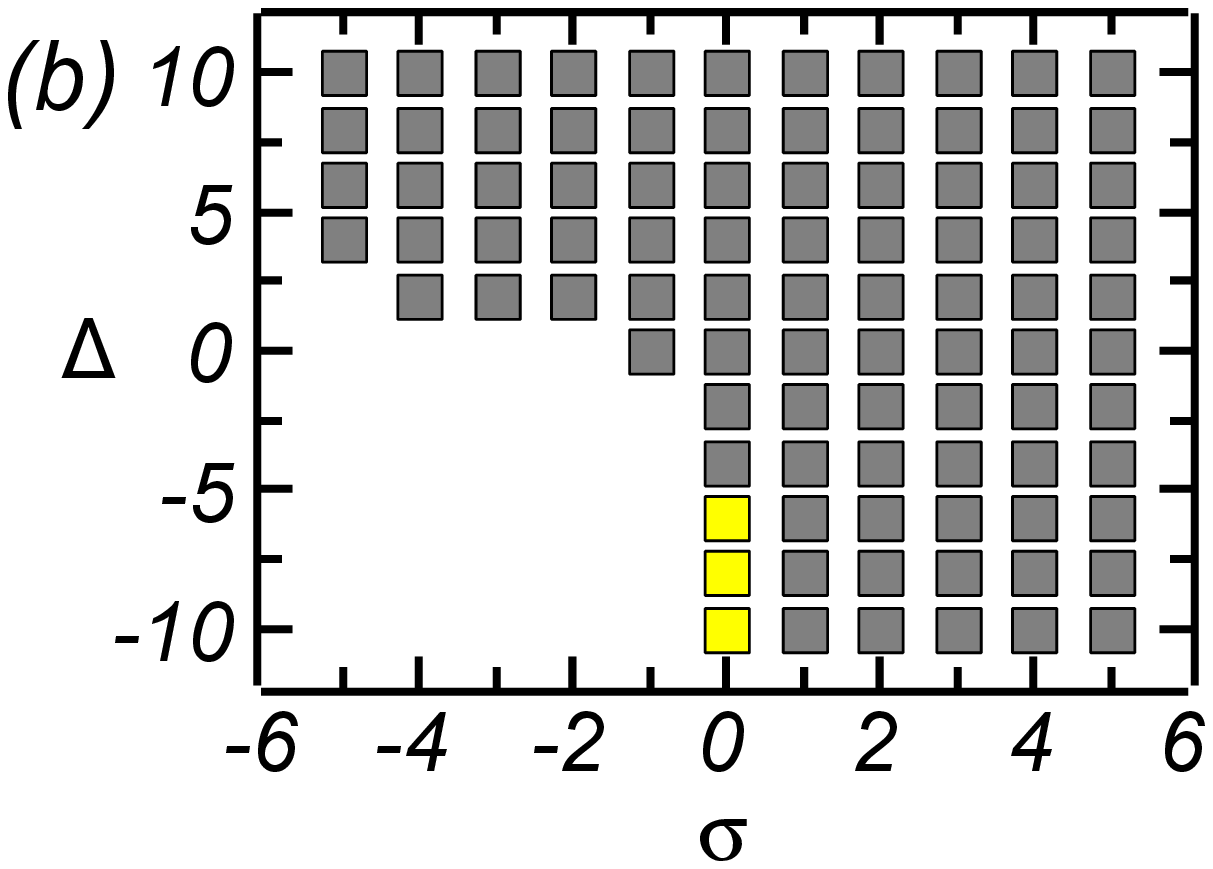}
\caption{(Color online) The stability area in the plane of ($\Delta $, $%
\protect\sigma $) for vortex solutions numerically generated by real-time
simulations of Eq. (\protect\ref{LuLe}) with vortex pump (\protect\ref{lin}%
). Other parameters are $\Omega =2$, $\protect\gamma =1$, and (a) $E_{0}=1$
or (b) $E_{0}=2$. Simple stable vortices are found in the gray area, while
the yellow one represents stable modes with the spiral phase structure which
features a full turn, and a multi-ring radial structure, see a typical
example in Fig. \protect\ref{strong spiral}. No stable vortices were found
in the white area.}
\label{VF1}
\end{figure}

\begin{figure}[tb]
\centering \includegraphics[width=0.48\columnwidth]{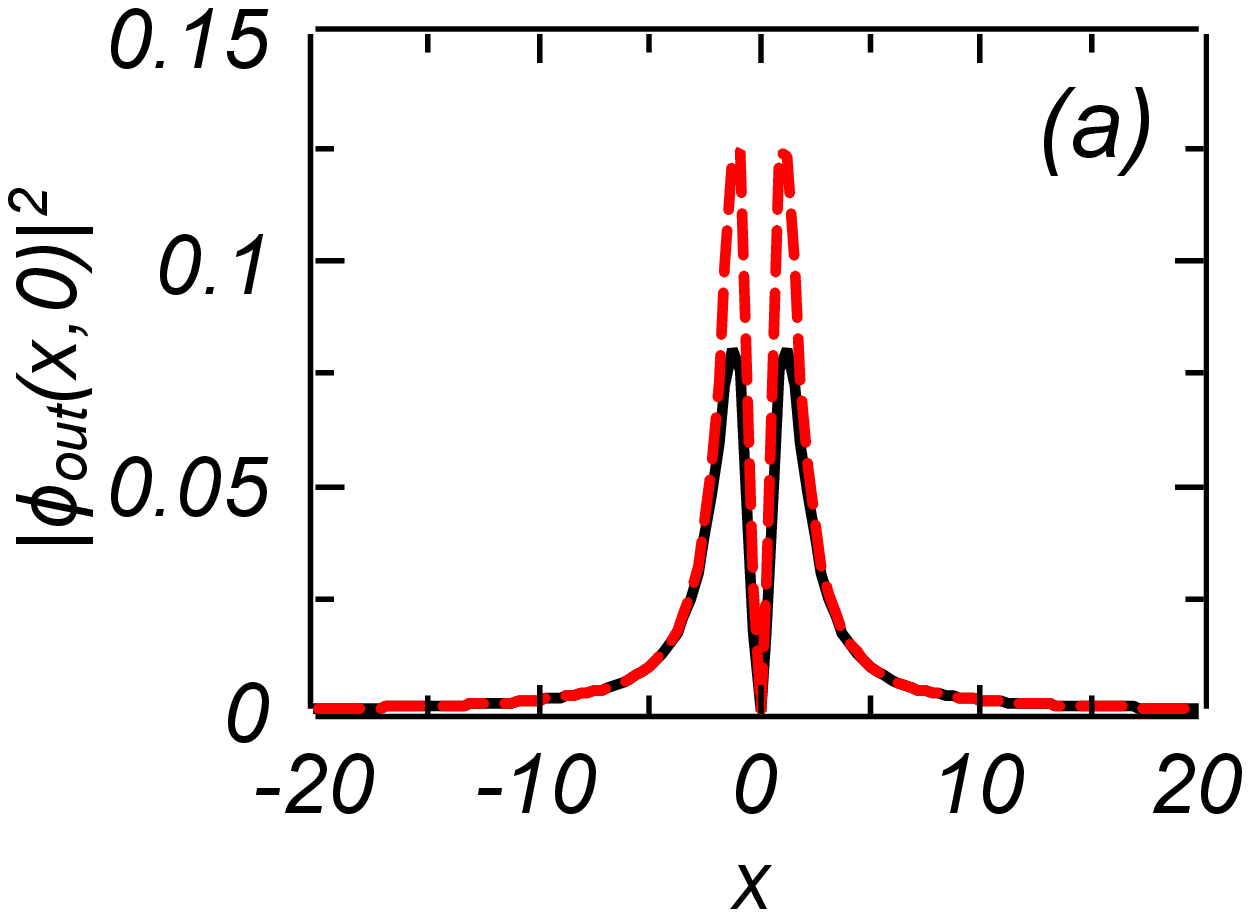} %
\includegraphics[width=0.48\columnwidth]{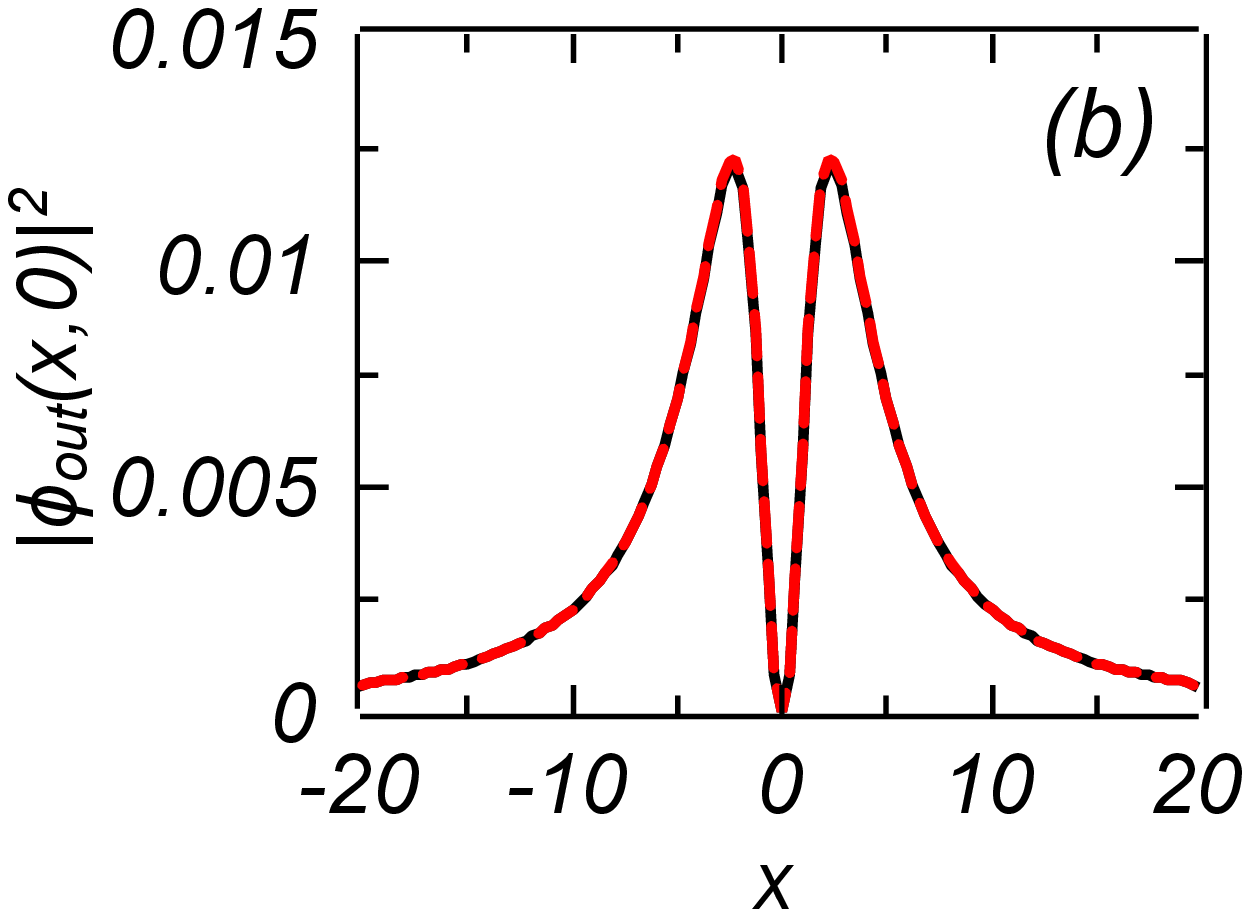}
\caption{(Color online) Output profiles $|\protect\phi _{\mathrm{out}%
}(x,0)|^{2}$ of stable ring-shaped vortices, produced by real-time
integration of Eq. (\protect\ref{LuLe}) with pump profile (\protect\ref{lin}%
), for two different values of $|\protect\sigma |$ (the absolute value of
the nonlinearity coefficient). The radial shapes obtained with the
self-defocusing ($\protect\sigma =5$) and focusing ($\protect\sigma =-5$)
nonlinearities are displayed by solid black and dashed red lines,
respectively, for $\Delta =0$ (a) and $\Delta =10$ (b). Other parameters are
$\protect\gamma =1$, $\Omega =2$ and $E_{0}=1$.}
\label{VF2}
\end{figure}

In the previous sections, we considered uniform pump field $E$, which
generates fundamental modes without vorticity. Here we explore the confined
LL model with space-dependent pump carrying the vorticity. It is represented
by the driving term
\begin{equation}
E=E_{0}re^{i\theta }  \label{lin}
\end{equation}%
in Eq. (\ref{LuLe}), where $\theta $ is the angular coordinate and $E_{0}=%
\mathrm{const}$. This term naturally corresponds to the pump supplied by a
vortex laser beam (with vorticity $1$) \cite{vortex beam}. In the case of
multiple vorticity $m>1$ (which will be considered elsewhere), Eq. (\ref{lin}%
) is replaced by $E=E_{0}r^{m}e^{im\theta }$.

Patterns supported by the vortex pump correspond to factorized solutions of
the stationary version of Eq. (\ref{LuLe}), taken as%
\begin{equation}
\phi \left( r,\theta \right) =e^{i\theta }A(r),  \label{A}
\end{equation}%
with complex amplitude $A$ satisfying the following radial equation:
\begin{gather}
\left[ \frac{1}{2}\left( \frac{d^{2}}{dr^{2}}+\frac{1}{r}\frac{d}{dr}-\frac{1%
}{r^{2}}\right) -\Delta +i\gamma -\frac{\Omega ^{2}}{2}r^{2}-\sigma |A|^{2}%
\right] A=  \notag \\
E_{0}r\;.  \label{ODE}
\end{gather}%
As an analytical approximation, the TFA for vortex solitons may be applied
here, cf. Ref. \cite{TFA-vortex}. In the general case, the TFA implies
dropping the derivatives in the radial equation, which leads to a complex
cubic equation for $A$, cf. Eq. (\ref{TF}), under the conditions $\sigma >0$
(self-defocusing) and $\Delta >0$ (positive mismatch):%
\begin{equation}
\left[ \Delta -i\gamma +\frac{1}{2}\left( \frac{1}{r^{2}}+\Omega
^{2}r^{2}\right) +\sigma |A|^{2}\right] A=-E_{0}r\;.  \label{TF-vortex}
\end{equation}

Equation (\ref{TF-vortex}), as well as its counterpart (\ref{TF})\ for the
zero-vorticity states, strongly simplifies in the limit of large $\Delta >0$%
, when both the imaginary and and nonlinear terms may be neglected:%
\begin{equation}
A(r)=-E_{0}r\left[ \Delta +\frac{1}{2}\left( \frac{1}{r^{2}}+\Omega
^{2}r^{2}\right) \right] ^{-1}.  \label{A(r)}
\end{equation}%
In particular, the simplest approximation provided by Eq. (\ref{A(r)}) makes
it possible to easily predict the radial location of maximal intensity in
the ring-shaped vortex mode:%
\begin{equation}
r_{\max }^{2}=\left( \sqrt{\Delta ^{2}+3\Omega ^{2}}+\Delta \right) /\Omega
^{2}.  \label{max}
\end{equation}%
Comparison of values given by Eq. (\ref{max}) with their counterparts
extracted from numerically found vortex-ring shapes, which are displayed
below in Figs. \ref{VF2}(a) and (b) for $\Delta \geq 0$, demonstrates that
the analytically predicted values are smaller than the numerical
counterparts by $11\%$ for $\Delta =0$, and by $6\%$ for $\Delta =10$.
Naturally, the TFA provides better accuracy for large $\Delta ,$ but even
for $\Delta =0$ the prediction is reasonable. Furthermore, Eq. (\ref{A(r)})
predicts a virtually exact largest intensity, $\left\vert A(r=r_{\max
})\right\vert ^{2}$, for the small-amplitude mode displayed in Fig. \ref{VF2}%
(b).

\subsection{Numerical results}

Equation (\ref{LuLe}) with vortex pump profile (\ref{lin}) was numerically
solved with zero input. This simulation scenario is appropriate, as vortex
states, when they are stable, are sufficiently strong attractors to draw
solutions developing from the zero input.

\begin{figure}[tb]
\includegraphics[width=0.48\columnwidth]{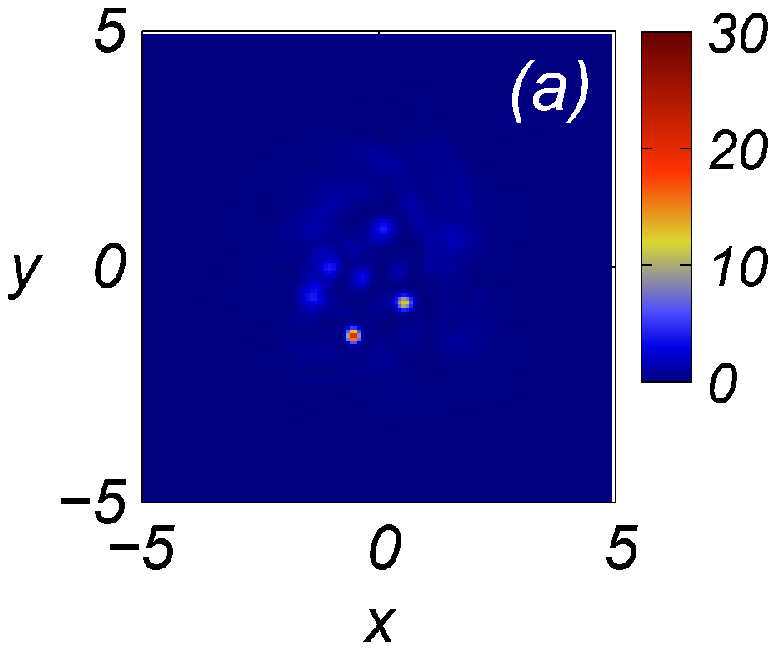} %
\includegraphics[width=0.48\columnwidth]{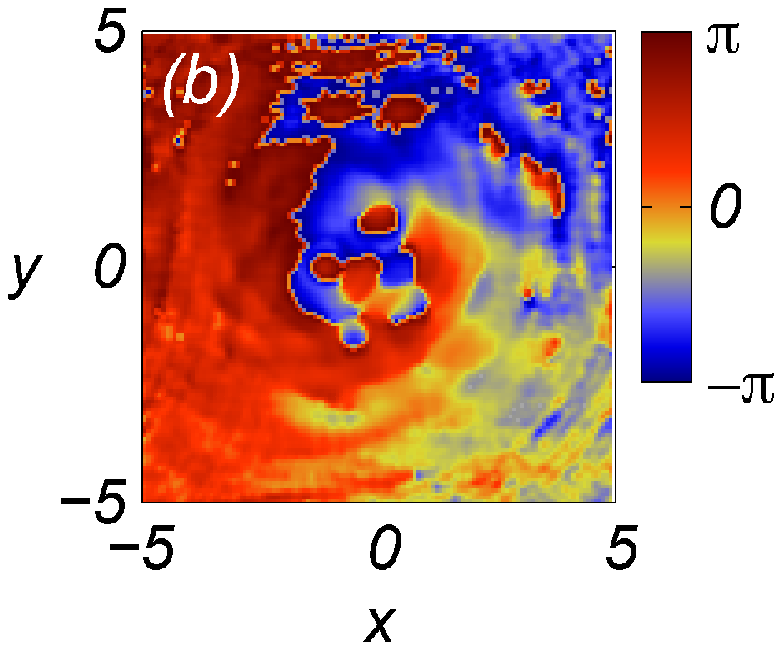}
\caption{(Color online) (a) Local-intensity $|\protect\phi \left( x,y\right)
|^{2}$ and (b) phase profiles of an unstable pattern, produced by the
simulations of Eq. (\protect\ref{LuLe}) with vortex pump (\protect\ref{lin})
and the strong self-focusing ($\protect\sigma =-5$) at $t=20$. Other
parameters are $\Delta =-8$, $\protect\gamma =1$, $\Omega =2$, and $E_{0}=1$%
. }
\label{VF3}
\end{figure}

The results, produced by systematic real-time simulations, are summarized in
Fig. \ref{VF3} for $\Omega =2$ in Eq. (\ref{LuLe}) and $E_{0}=1$ or $2$ in
Eq. (\ref{lin}). The figure displays stability areas for the vortex modes in
the plane of free control parameters ($\Delta $, $\sigma )$ (the mismatch
and nonlinearity strength). It is worthy to note that the stability domain
for the self-focusing nonlinearity ($\sigma <0$) is essentially larger than
in the diagram for the fundamental (zero-vorticity) modes, which is
displayed, also for $\Omega =2$, in Fig. \ref{NER}. This fact may be
naturally explained by the fact that the vanishing of the vortex drive (\ref%
{lin}) at $r\rightarrow 0$, in the combination with the intrinsic structure
of the vortex states, makes the central area of the pattern nearly
``empty", thus preventing the onset of the modulational
instability in it.

In the gray areas in Fig. \ref{VF1}, the stable vortex modes have a simple
ring-shaped structure, with typical radial profiles shown in Fig. \ref{VF2}.
In the case of zero mismatch, $\Delta =0$ [Fig. \ref{VF1}(a)], the vortex
state naturally acquires a higher amplitude under the action of the
self-focusing. On the other hand, in the case of large positive mismatch
[Fig. \ref{VF1}(b)], the small amplitude is virtually the same under the
action of the focusing and defocusing, which is explained, as mentioned
above, by the TFA that reduces to Eq. (\ref{A(r)}).

In unstable (white) areas in Fig. \ref{VF1}, direct simulations lead to
quick fragmentation of vortically driven patterns into small spots, that
feature a trend to developing the above-mentioned critical collapse \cite%
{collapse}. A typical example of the unstable evolution is displayed in Fig. %
\ref{VF3}.

\begin{figure}[tb]
\includegraphics[width=0.48\columnwidth]{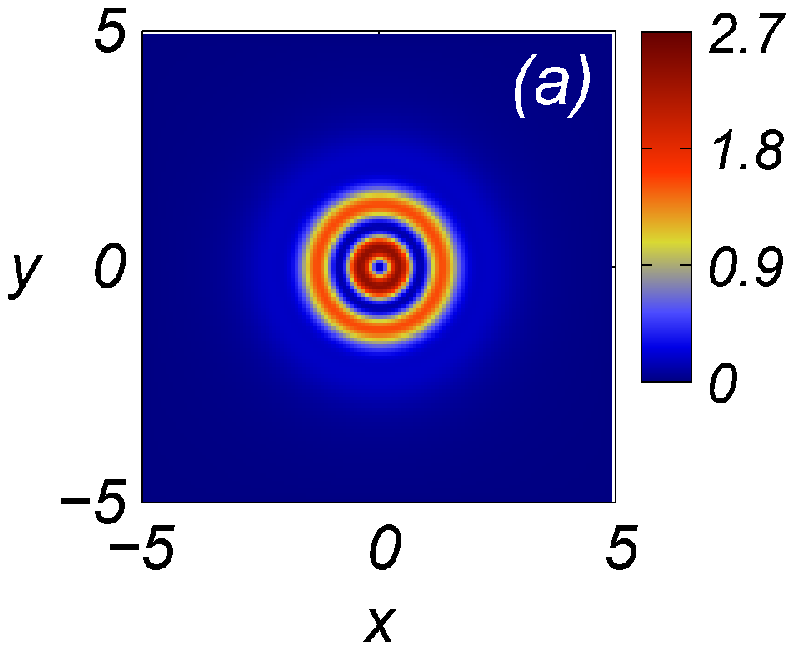}%
\includegraphics[width=0.48\columnwidth]{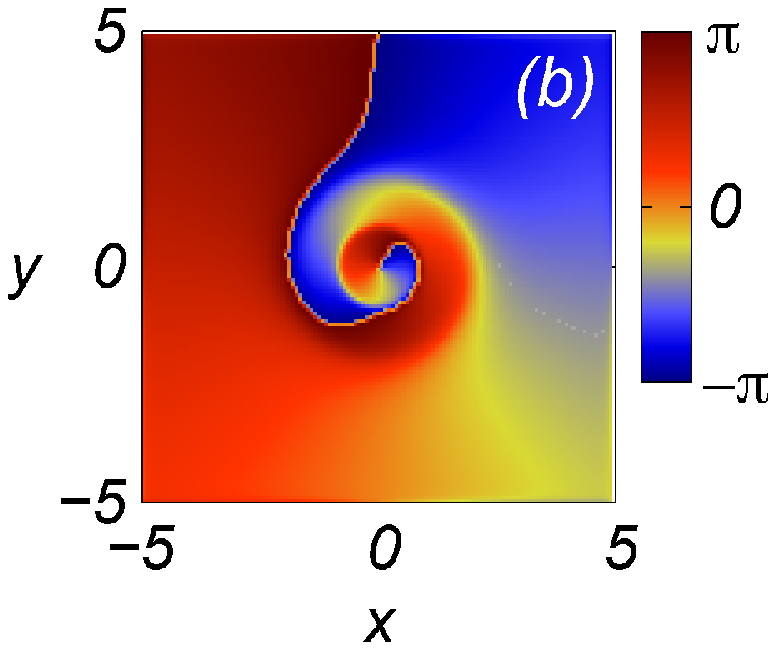} %
\includegraphics[width=0.48\columnwidth]{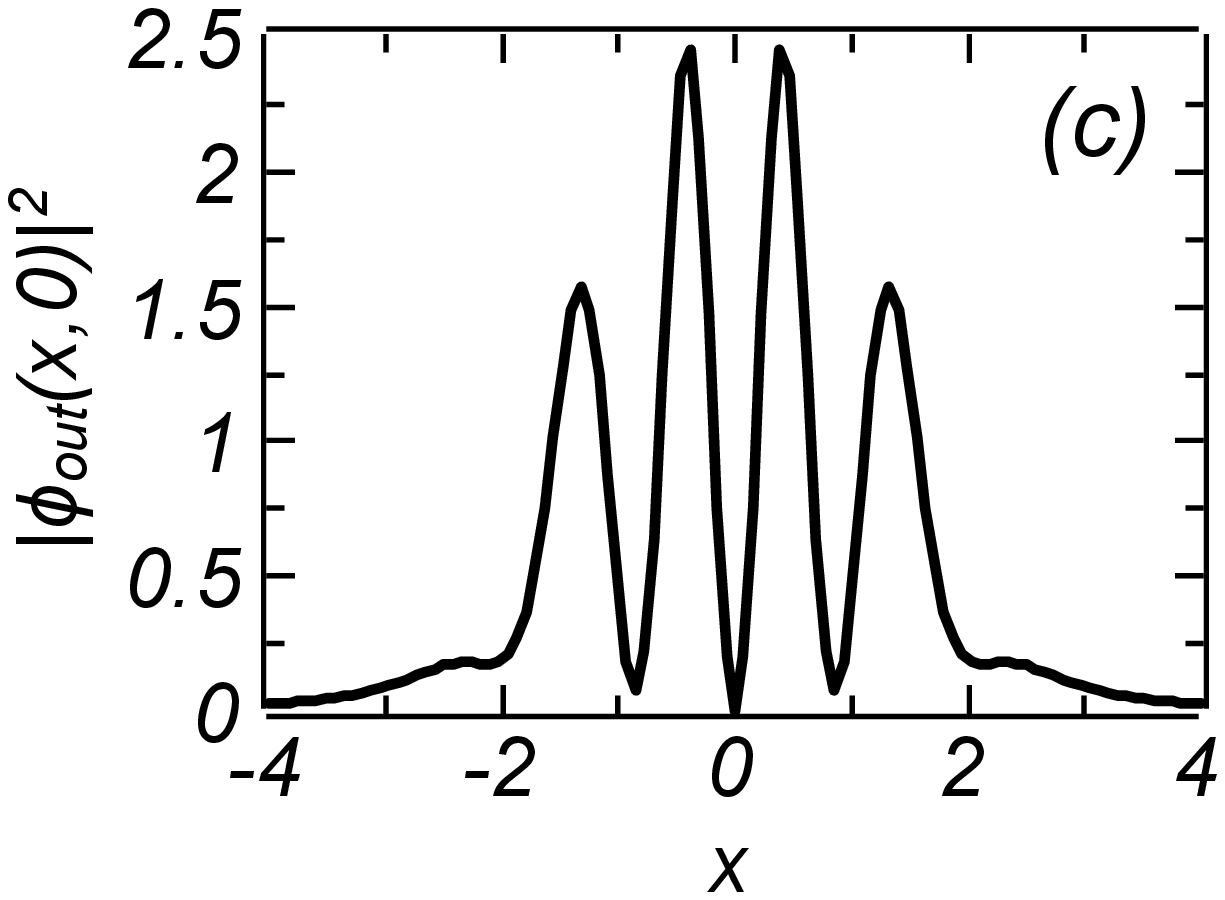}
\caption{(Color online) A stable multi-ring vortex with the spiral phase
field. Panels (a), (b), and (c) display, respectively, the 2D
local-intensity pattern, phase field, and the radial structure. Parameters
are the same as in Fig. \protect\ref{VF3}, except for a weaker self-focusing
strength, $\protect\sigma =-1$.}
\label{strong spiral}
\end{figure}

More sophisticated stable vortex profiles are observed in yellow areas in
Fig. \ref{VF3}. They are characterized by a multi-ring radial structure, and
a spiral shape of the vorticity-carrying phase distribution, as shown in
Fig. \ref{strong spiral}. The yellow areas are defined as those in\ which
the spiral phase field performs a full turn by $360$ degrees, as can be seen
in Fig. \ref{strong spiral}(b). Note that this area exists for both the
focusing and defocusing signs of the nonlinearity in Fig. \ref{VF3}(a), and
solely for zero nonlinearity in Fig. \ref{VF3}(b), which corresponds to the
stronger pump.

The spiral shape of the phase pattern is explained by the fact that radial
amplitude $A(r)$ in solution (\ref{A}) is a complex function, as is
explicitly shown, in particular, by Eqs. (\ref{infi}) and (\ref{TF-vortex}).
The spirality of vortices is a well-known feature of 2D complex
Ginzburg-Landau equations \cite{GL}. However, unlike the present situation,
the spirality is not usually related to a multi-ring radial structure.
Patterns with multi-ring shapes usually exist as excited states, on top of
stable ground states in the same models, being unstable to azimuthal
perturbations \cite{Javid}. For this reason, the stability of complex modes,
like the one displayed in Fig. \ref{strong spiral}, is a noteworthy finding.

\begin{figure}[tb]
\includegraphics[width=0.48\columnwidth]{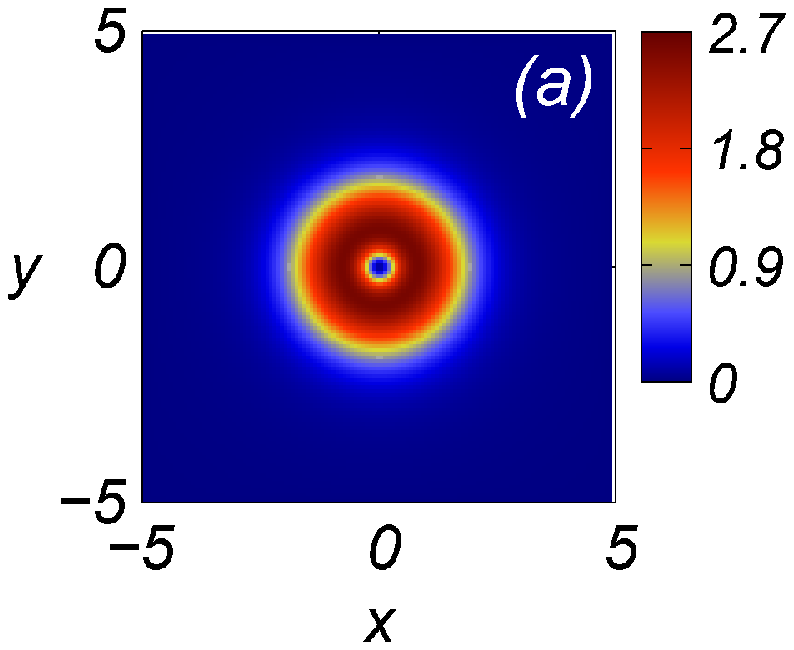} %
\includegraphics[width=0.48\columnwidth]{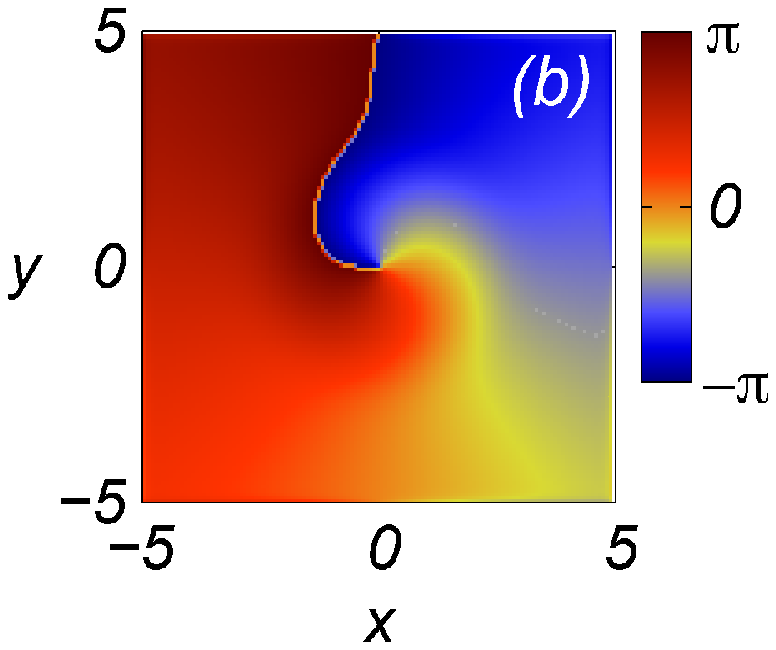} %
\includegraphics[width=0.48\columnwidth]{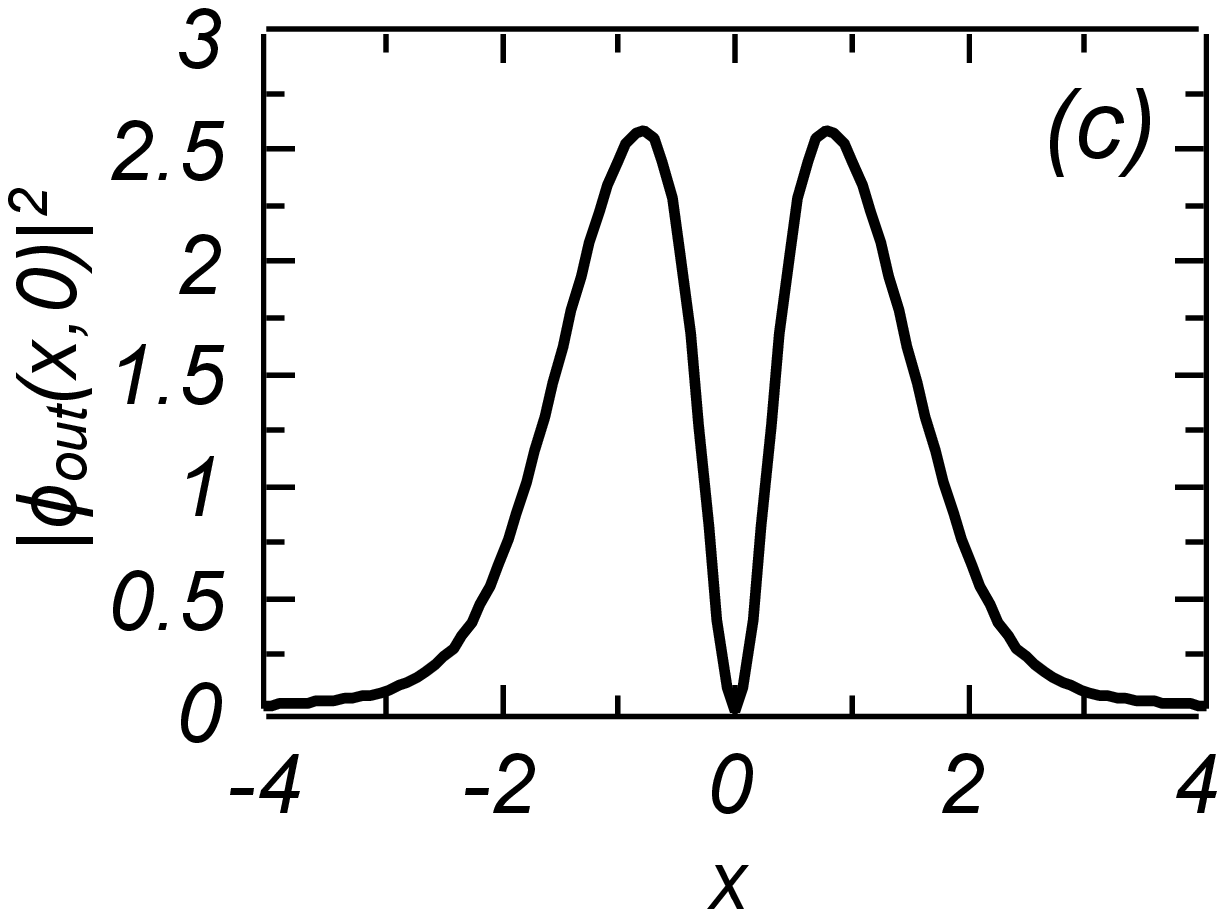}
\caption{(Color online) The same as in Fig. \protect\ref{VF3}, but with the
self-defocusing sigh of the nonlinearity, $\protect\sigma =2$.}
\label{weak spiral}
\end{figure}

Lastly, a typical example of a stable vortex at the boundary between the
simple (non-spiral) and complex (spiral-shaped) ones is presented in Fig. %
\ref{weak spiral}. It features emerging spirality in the phase field, but
the radial structure keeps the single-ring shape.

\section{Conclusion}

We have introduced the 2D model based on the LL\ (Lugiato-Lefever) equation
with confinement imposed by the harmonic-oscillator trap. In spite of the
action of the uniform pump, the confinement creates well localized patterns,
which may be used for the creation of robust small-area pixels in
applications. The VA (variational approximation), based on a novel
fractional ansatz, as well as a simple TFA (Thomas-Fermi approximation),
were elaborated to describe the fundamental (zero-vorticity) confined modes.
The VA effectively reduces the 2D LL equation to the zero-dimensional
version. The VA is additionally enhanced by taking into regard the balance
condition for the integral power. The comparison with the full numerical
analysis has demonstrated that the VA provides qualitatively accurate
predictions, which are also quantitatively accurate, in some areas of the
parameter space. The systematic numerical analysis has produced overall
stability areas for the confined pattern in the underlying parameter space,
which demonstrate that the patterns tend to be less stable and more stable
under the action of the self-focusing and defocusing nonlinearity,
respectively (although very strong self-defocusing causes fragmentation of
the patterns). The increase of the confinement strength leads to shrinkage
of the stability area, although it does not make all the states unstable. On
the other hand, large positive values of the cavity's detuning tends to
expand the region of the stability in the parameter space.

We have also explored vortex solitons (which may be used to realize vortical
pixels in microcavities) supported by the pump with embedded vorticity. In
this case, the simple TFA provides a qualitatively correct description, and
systematically collected numerical results reveal a remarkably large
stability area in the parameter space, for both the self-defocusing and
focusing signs of the nonlinearity. In addition to simple vortices, stable
complex ones, featuring the multi-ring radial structure and the spiral phase
field, have been found too. As an extension of the present work, a
challenging issue is to look for confined states with multiple embedded
vorticity.

A summary of authors' contributions to the work: the numerical part has been
carried out by W.B.C. Analytical considerations were chiefly performed by
B.A.M. and L.S. All the authors have contributed to drafting the text of the
paper.

\begin{acknowledgments}
WBC acknowledges the financial support from the Brazilian agencies CNPq
(grant \#458889/2014-8) and the National Institute of Science and Technology
(INCT) for Quantum Information. LS acknowledges for partial support the 2016 BIRD project
``Superfluid properties of Fermi gases in optical potentials''
of the University of Padova. The work of B.A.M. is supported, in part, by grant No. 2015616 from
the joint program in physics between NSF (US) and
Binational (US-Israel) Science Foundation.
\end{acknowledgments}


\begin{thebibliography}{99}

\bibitem{LL} L. A. Lugiato and R. Lefever, \emph{Spatial dissipative
structures in passive optical systems}, Phys. Rev. Lett. \textbf{58},
2209-2211 (1987).

\bibitem{patterns-early} G.-L. Oppo, M. Brambilla, and L. A. Lugiato, \emph{%
Formation and evolution of roll patterns in optical parametric oscillators},
Phys. Rev. A \textbf{49}, 2028-2032 (1994).

\bibitem{pixel} M. Brambilla, L. A. Lugiato, F. Prati, L. Spinelli, and W.
J. Firth, \emph{Spatial soliton pixels in semiconductor devices}, Phys. Rev.
Lett. \textbf{79}, 2042 (1997).

\bibitem{patterns-2} L. Gelens, D. Gomila, G. Van der Sande, J. Danckaert,
P. Colet, and M. A. Mat\'{\i}as, \emph{Dynamical instabilities of
dissipative solitons in nonlinear optical cavities with nonlocal materials},
Phys. Rev. A \textbf{77}, 033841 (2008); T. Miyaji, I. Ohnishi,and Y.
Tsutsumi, \emph{Bifurcation analysis to the Lugiato-Lefever equation in one
space dimension}, Physica D 239, 2066-2083 (2010); K. Panajotov, D. Puzyrev,
A. G. Vladimirov, S. V. Gurevich, and M. Tlidi, \textit{Impact of
time-delayed feedback on spatiotemporal dynamics in the Lugiato-Lefever model%
}, Phys. Rev. A \textbf{93}, 043835 (2016).

\bibitem{Kestas} G. J. de Valc\'{a}rcel and K. Staliunas, \emph{%
Phase-bistable Kerr cavity solitons and patterns}, Phys. Rev. A \textbf{87},
043802 (2013).

\bibitem{patterns} P. Parra-Rivas, D. Gomila, M. A. Mat\'{\i}as, S. Coen,
and L. Gelens, \emph{Dynamics of localized and patterned structures in the
Lugiato-Lefever equation determine the stability and shape of optical
frequency combs}, Phys. Rev. A \textbf{89}, 043813 (2014); C. Godey, I. V.
Balakireva, A. Coillet, Aurelien, and Y. K. Chembo, \emph{Stability analysis
of the spatiotemporal Lugiato-Lefever model for Kerr optical frequency combs
in the anomalous and normal dispersion regimes}, \textit{ibid}. \textbf{89},
063814 (2014); T. Hansson and S. Wabnitz, \emph{Frequency comb generation
beyond the Lugiato-Lefever equation: multi-stability and super cavity
solitons}, J. Opt. Soc. Am. B \textbf{32}, 1259-1266 (2015); F. Copie, M.
Conforti, A. Kudlinski, and A. Mussot, and S. Trillo, \emph{Competing Turing
and Faraday instabilities in longitudinally modulated passive resonators},
Phys. Rev. Lett. \textbf{116}, 143901 (2016); P. Parra-Rivas, E. Knobloch,
D. Gomila, and L. Gelens, \emph{Dark solitons in the Lugiato-Lefever
equation with normal dispersion}, Phys. Rev. A \textbf{93}, 063839 (2016).

\bibitem{NJP} J. K. Jang, M. Erkintalo, K. Luo, G.-L. Oppo, S. Coen, and S.
G. Murdoch, \emph{Controlled merging and annihilation of localised
dissipative structures in an AC-driven damped nonlinear Schr\"{o}dinger
system}, New J. Phys. \textbf{18}, 0336034 (2016).

\bibitem{VA-LL} A. B. Matsko and L. Maleki, \emph{On timing jitter of mode
locked Kerr frequency combs}, Opt. Exp. \textbf{21}, 28862 (2013).

\bibitem{TFA} L. P. Pitaevskii and A. Stringari, \emph{Bose-Einstein
Condensation} (Clarendon Press, Oxford, 2003).

\bibitem{IT} E. Cerboneschi, R. Mannella, E. Arimondo, and L. Salasnich,
\textit{Oscillation frequencies for a Bose condensate in a triaxial magnetic
trap}, Phys. Lett. A \textbf{249}, 495-5000 (1998); M. L. Chiofalo, S.
Succi, and M. P. Tosi,\emph{\ Ground state of trapped interacting
Bose-Einstein condensates by an explicit imaginary-time algorithm}, Phys.
Rev. E \textbf{62}, 7438 (2000); X. Antoine, W. Baoc, and C. Besse, \emph{%
Computational methods for the dynamics of the nonlinear Schr\"{o}%
dinger/Gross--Pitaevskii equations}, Comp. Phys. Commun. \textbf{184}, 2621
(2013).

\bibitem{Eaton_Octave} J. W. Eaton, D. Bateman, and S. Hauberg, \emph{GNU
Octave Manual - Version 3} (Network Theory Ltd., UK, 2008).

\bibitem{Yang_10} J. Yang, \emph{Nonlinear waves in integrable and
nonintegrable systems} (SIAM, Philadelphia, USA, 2010).

\bibitem{MI} S. Coen, H. G. Randle, S. Thibaut, and M. Erkinalo, \emph{%
Modeling of octave-spanning Kerr frequency combs using a generalized
mean-field Lugiato-Lefever model}, Opt. Lett. \textbf{38}, 37 (2013); C.
Godey, I. V. Balakireva, A. Coillet, and Y. K. Chembo, \emph{Stability
analysis of the spatiotemporal Lugiato-Lefever model for Kerr optical
frequency combs in the anomalous and normal dispersion regimes}, Phys. Rev.
A \textbf{89}, 063814 (2014); T. Hansson and S. Wabnitz, \emph{Dynamics of
microresonator frequency comb generation: models and stability},
Nanophotonics \textbf{5}, 231 (2016).

\bibitem{collapse} L. Berg\'{e}, \emph{Wave collapse in physics: Principles
and applications to light and plasma waves}, Phys. Rep. \textbf{303}, 259
(1998); G. Fibich, \emph{The Nonlinear Schr\"{o}dinger Equation: Singular
Solutions and Optical Collapse} (Springer: Heidelberg, 2015).

\bibitem{vortex beam} L. Allen, M. W. Beijersbergen, R. J. C. Spreeuw, and
J. P. Woerdman, \emph{Orbital angular momentum of light and the
transformation of Laguerre-Gaussian laser modes}, Phys. Rev. A \textbf{45},
8185 (1992); I. V. Basistiy, V. Yu. Bazhenov, M. S. Soskin, and M. V.
Vasnetsov, \emph{Optics of light beams with screw dislocations}, Opt.
Commun. \textbf{103}, 422 (1993); S. Franke-Arnold, L. Allen, and M.
Padgett, \emph{Advances in optical angular momentum}, Laser \& Photon. Rev.
\textbf{2}, 299 (2008).

\bibitem{TFA-vortex} L. Salasnich, A. Parola, and L. Reatto, \textit{Bosons
in a toroidal trap: Ground state and vortices}, Phys. Rev. A \textbf{59},
2990 (1999); A. L. Fetter, \emph{Rotating trapped Bose-Einstein condensates}%
, Rev. Mod. Phys.\textbf{\ 81}, 647 (2009); L. Salasnich and B.A. Malomed,
\textit{Solitons and solitary vortices in pancake-shaped Bose-Einstein
condensates}, Phys. Rev. A \textbf{79}, 053620 (2009); R. Driben, Y. V.
Kartashov, B. A. Malomed, T. Meier, and L. Torner, \emph{Soliton gyroscopes
in media with spatially growing repulsive nonlinearity}, Phys. Rev. Lett.
\textbf{112}, 020404 (2014); J. Qin, G. Dong, and B. A. Malomed, \emph{%
Stable giant vortex annuli in microwave-coupled atomic condensates}, Phys.
Rev. A \textbf{94}, 053611 (2016).

\bibitem{GL} T. Bohr, G. Huber, and E. Ott, \emph{The structure of
spiral-domain patterns and shocks in the 2D complex Ginzburg-Landau equation}%
, Physica D \textbf{106}, 95 (1997); M. Gabbay, E. Ott, and P. N. Guzdar,
\emph{The dynamics of scroll wave filaments in the complex Ginzburg-Landau
equation}, \textit{ibid}. \textbf{118}, 371 (1998); L.-C. Crasovan, B. A.
Malomed, and D. Mihalache, \emph{Stable vortex solitons in the
two-dimensional Ginzburg-Landau equation}, Phys. Rev. E \textbf{63}, 016605
(2001); D. Mihalache, D. Mazilu, F. Lederer, Y. V. Kartashov, L.-C.
Crasovan, L. Torner, and B. A. Malomed, \emph{Stable vortex tori in the
three-dimensional cubic-quintic Ginzburg-Landau equation}, Phys. Rev. Lett.
\textbf{97}, 073904 (2006).

\bibitem{Javid} J. Atai, Y. J. Chen, and J. M. Soto-Crespo, \emph{Stability
of 3-dimensional self-trapped beams with a dark spot surrounded by bright
rings of varying intensity}, Phys. Rev. A \textbf{49}, R3170 (1994); A.
Dubietis,G. Tamosauskas, G. Fibich, and B. Ilan, \emph{Multiple
filamentation induced by input-beam ellipticity}, Opt. Lett. \textbf{29},
1126 (2004).

\end{thebibliography}
\end{document}